\documentclass[trackchanges, twocolumn, resetfootnote, times, twocolappendix]{aastex702}

\newcommand{\sersic}{S\'ersic }
\newcommand{\pysersic}{\texttt{pysersic} }
\newcommand{\pysercomma}{\texttt{pysersic}}

\newcommand{\bagpipes}{\texttt{BAGPIPES} }
\newcommand{\pixedfit}{\texttt{piXedfit} }
\newcommand{\pixcomma}{\texttt{piXedfit}}
\newcommand{\bgcomma}{\texttt{BAGPIPES}}
\DeclareRobustCommand{\AA}{\mbox{\fontencoding{T1}\selectfont\symbol{197}}}
\newcommand{\id}{\texttt{ID} }

\shorttitle{Size--Mass Relation \& Quenching at $z\sim 2$--$3$} 
\shortauthors{Prasal et al.}

\begin{document}

\title{The Size-Mass Relation and Inside-Out Quenching of Massive Quiescent Galaxies at $2<z<3$ with JWST PRIMER and MINERVA}

\author[orcid=0009-0007-4725-1455,sname=Prasal,gname=Vaidik]{Vaidik Prasal}
\affiliation{Indian Institute of Science Education and Research Pune, Dr. Homi Bhabha Road, Pashan, Pune 411008, India}\affiliation{National Centre for Radio Astrophysics – Tata Institute of Fundamental Research, Post Bag 3, Ganeshkhind, Pune 411007, India}
\email{vaidik.prasal@alumni.iiserpune.ac.in}

\author[orcid=0000-0002-1345-7371,sname=Wadadekar,gname=Yogesh]{Yogesh Wadadekar}
\affiliation{National Centre for Radio Astrophysics – Tata Institute of Fundamental Research, Post Bag 3, Ganeshkhind, Pune 411007, India}
\email{yogesh@ncra.tifr.res.in}

\author[orcid=0009-0002-2173-0953,sname=Biswas,gname=Pralay]{Pralay Biswas}
\affiliation{National Centre for Radio Astrophysics – Tata Institute of Fundamental Research, Post Bag 3, Ganeshkhind, Pune 411007, India}
\email{pbiswas@ncra.tifr.res.in}

\email[show]{vaidik.prasal@alumni.iiserpune.ac.in}

\begin{abstract}
We investigate the morphology, size--mass relations, and spatially resolved star formation histories of 26 massive quiescent galaxies at cosmic noon ($2 < z < 3$) in the Ultra-Deep Survey field using deep JWST NIRCam imaging from the PRIMER and MINERVA surveys. Combining 16 broadband and medium-band filters with empirical point spread function matching, we perform single-S\'ersic modeling and bulge--disk decompositions with \pysersic across six filters (F115W to F444W), alongside Bayesian size--mass fitting with \texttt{dynesty}. In the rest-frame optical (F200W), the size--mass relation exhibits a steep slope of $\alpha = 0.59 \pm 0.12$ and effective radius of $A = 0.83_{-0.07}^{+0.08}\ \mathrm{kpc}$ at $M_* = 5 \times 10^{10}\ M_\odot$. In the rest-frame near-infrared (F356W), we derive $\alpha = 0.56 \pm 0.13$ and a more compact radius of $A = 0.70_{-0.06}^{+0.07}\ \mathrm{kpc}$. Galaxy sizes shrink by $\approx 22\%$ from F115W to F444W, reflecting negative color gradients and centrally concentrated mass profiles. Two-thirds of the sample are bulge-dominated ($B/T \ge 0.5$), whereas one-third are disk-dominated. Central bulges are ultra-compact ($R_{e,\mathrm{Bulge}} \approx 0.49\ \mathrm{kpc}$) with negligible mass dependence, whereas outer disks extend to $R_{e,\mathrm{Disk}} \approx 1.4\text{--}1.5\ \mathrm{kpc}$. Resolved spectral energy distribution modeling with \pixedfit demonstrates positive radial specific star formation rate gradients and inner regions that formed at least $\approx 0.7\ \mathrm{Gyr}$ earlier and quenched faster than their outskirts. These results confirm inside-out quenching, consistent with dissipative gas compaction assembling the dense core, active galactic nucleus feedback halting central star formation, and the compact bulge dynamically stabilizing the surviving disk against fragmentation.
\end{abstract}


\section{Introduction} \label{sec:intro}

Massive galaxies at cosmic noon ($2 < z < 3$) are critical for understanding how early galactic structures assemble and why star formation ceases \citep{Cimatti2004, Glazebrook2004}. A defining feature of these early quiescent systems is their extreme compactness. They built up large stellar masses ($M_* \gtrsim 10^{10}\ M_\odot$) and halted star formation within the first few billion years after the Big Bang, during the peak epoch of cosmic star formation \citep{Madau2014}. These galaxies exhibit very compact structures, with typical half-light radii of $R_e \sim 0.6$--$1.5\ \mathrm{kpc}$, which are a factor of four to six smaller than local early-type galaxies of comparable stellar mass \citep[][see also \citealt{Haryana_2025}]{Buitrago2008, vanDokkum2008, vanderWel2014, Mowla2019b, Martorano2024}. Consequently, their mean stellar mass surface densities within the effective radius are approximately two orders of magnitude higher than those of their low-redshift counterparts \citep{Buitrago2008, vanDokkum2008}. Understanding the structural scaling relations of these compact galaxies and the physical mechanisms that rapidly halt star formation in their cores remains a foundational goal of extragalactic astronomy.

This structural evolution is parameterized across cosmic time by the galaxy size--mass relation ($R_e \propto M_*^\alpha$; \citealt{Shen2003, vanderWel2014, Mowla2019b}). While early studies often modeled this relation as a single power law, recent wide-field surveys reveal that the size--mass relation flattens below a pivot mass of $\log_{10}(M_*/M_\odot) \sim 10.3$ at lower redshifts \citep[e.g.,][]{Mowla2019a, Kawinwanichakij2021, Nedkova2021, Cutler2022}, suggesting distinct structural assembly pathways for low-mass and high-mass quiescent galaxies. Furthermore, rest-frame optical sizes can be significantly biased by radial color gradients and outshining from young stars \citep{Suess2019, Suess2022}; observing in the rest-frame near-infrared ($\lambda_{\mathrm{rest}} \gtrsim 1.0\ \mu\mathrm{m}$) with the James Webb Space Telescope (JWST) is therefore essential to trace the true distribution of stellar mass and establish unbiased structural scaling relations \citep{Martorano2024, Kawinwanichakij2026}.

This compact structure and dense core are intimately connected to the paradigm of inside-out quenching, where star formation ceases first in the central regions while the outskirts remain active \citep{Tacchella2015, Tacchella2016, Spilker2019}. This compact morphology can arise through gas-rich dissipative compaction (the ``blue nugget'' phase) that funnels gas into the galactic center to build a dense core prior to quenching \citep{Dekel2014, Zolotov2015, Barro2017}. Following core formation, star formation shuts down either through morphological quenching, where a dominant stellar bulge stabilizes the remaining disk against fragmentation \citep{Martig2009}, or through active galactic nucleus (AGN) feedback, where radiatively driven winds or kinetic jets heat or expel central gas reservoirs \citep{Croton2006, Fabian2012,Nelson2019, Nelson2021, Park2024, Scholtz2026}. Recent JWST spatially resolved studies directly confirm this inside-out quenching picture at $z \sim 2$--$3$ through positive radial specific star formation rate (sSFR) gradients and older central stellar ages \citep{Haryana_2025,Laishram2026,Prasal2026}.

Disentangling quenching pathways and scaling relations requires reliable stellar population modeling. Broadband-only spectral energy distribution (SED) fitting suffers from age–dust–metallicity degeneracies and cannot separate strong emission lines from the stellar continuum. As a result, broadband-only fits can significantly bias stellar mass and star formation rate (SFR) estimates \citep{Sarrouh2024, Martis2025}. Medium-band filters on the JWST Near-Infrared Camera (NIRCam) overcome these biases by providing quasi-spectroscopic continuum sampling across rest-frame optical wavelengths. By cleanly isolating the $4000\ \mathrm{\AA}$ and Balmer breaks, tracing continuum curvature, and separating line emission, medium bands reduce the photometric redshift scatter and catastrophic outlier rates by factors of $\sim 2$--$4$ and $\sim 2$--$3$, respectively, and the stellar mass scatter by a factor of $\sim 2$ \citep{Suess2024, Muzzin2025}.

In this paper, we investigate the size--mass relation, structural morphology, and inside-out quenching of 26 massive quiescent galaxies ($M_* > 10^{10}\ M_\odot$, $\mathrm{sSFR} < 0.2/t_{\mathrm{age}}$) at cosmic noon ($2 < z < 3$) in the Ultra-Deep Survey (UDS) field. We utilize coordinated JWST/NIRCam broadband and medium-band imaging from the PRIMER \citep{Donnan2024} and MINERVA \citep{Muzzin2025} Treasury surveys, assembling a 26-band dataset spanning the Hubble Space Telescope (HST), 16 NIRCam filters, and the Mid-Infrared Instrument (MIRI). Using \bgcomma\ \citep{Carnall2018}, empirical point-spread function (PSF) construction and homogenization to F444W \citep{Boucaud2016, Weaver2024}, and \pysercomma\ \citep{Pasha2023} modeling across six NIRCam broadband filters (F115W to F444W), we fit the size--mass relation above $\log_{10}(M_*/M_\odot) > 10.3$ using \texttt{dynesty} \citep{Speagle2020} with COSMOS2020 mass-function weighting \citep{Weaver2023, Kawinwanichakij2026}, and trace its wavelength dependence from rest-frame optical to near-infrared.

This paper is organized as follows. Section~\ref{sec:dataset} describes the PRIMER, MINERVA, and archival HST datasets. Section~\ref{sec:methods} details our SED modeling with \bgcomma, sample selection criteria, empirical PSF construction and homogenization, and morphological modeling with \pysercomma\ and \texttt{dynesty}. Section~\ref{sec:results} presents our structural morphology, multi-wavelength size--mass scaling relations, two-component decompositions, and spatially resolved stellar population radial profiles. Section~\ref{sec:discussion} discusses cosmic size evolution, the two-phase assembly paradigm, inside-out quenching via dissipative compaction, and the physical drivers of star formation cessation. Section~\ref{sec:conclusions} summarizes our primary conclusions. Throughout this work, we adopt the \texttt{Planck2018} cosmology with $H_0 = 67.66\ \mathrm{km\ s^{-1}\ Mpc^{-1}}$, $\Omega_m = 0.31$, and $\Omega_\Lambda = 0.69$ \citep{Planck2018}.

\section{Dataset} \label{sec:dataset}

We utilize image mosaics, weight maps, and photometric products from coordinated JWST surveys of the UDS field. This dataset primarily combines observations from the Public Release IMaging for Extragalactic Research (PRIMER; e.g., \citealt{Donnan2024}) JWST Cycle 1 Treasury program (PID-1837; PI: Dunlop) and the Medium-band Imaging with NIRCam to Explore ReVolutionary Astrophysics (MINERVA; \citealt{Muzzin2025}) JWST Cycle 4 Treasury program (PID-7814; PI: Muzzin, Co-PIs: Marchesini \& Suess). 

\subsection{PRIMER Survey}

The PRIMER survey observed the UDS field using MIRI in prime and NIRCam in parallel \citep{Muzzin2025}. The prime MIRI observations comprise 54 pointings in the F770W and F1800W filters with exposure times of $\sim 30$ minutes per pointing, covering $125\ \mathrm{arcmin}^2$ \citep[][Table~2]{Muzzin2025}. The corresponding parallel NIRCam mosaic covers $234\ \mathrm{arcmin}^2$ across 54 pointings through seven broadband filters (F090W, F115W, F150W, F200W, F277W, F356W, and F444W) and one medium-band filter (F410M), with exposure times of 14--28 minutes per filter \citep{Muzzin2025}. The NIRCam imaging reaches $5\sigma$ point-source depths of $\sim 27.6$--$28.3$ AB mag in NIRCam \citep{Donnan2024}). This provides the deep broadband baseline and spatial resolution required for structural and morphological analyses.

\subsection{MINERVA Survey}

To complement the PRIMER imaging, we incorporate data from the MINERVA survey, which provides comprehensive medium-band coverage over the same field \citep{Muzzin2025}. MINERVA duplicates the PRIMER pointing layout in UDS across 54 pointings, obtaining prime MIRI imaging in the F1280W and F1500W filters ($\sim 38$ minutes exposure per filter using a 3-point dither strategy, covering $125\ \mathrm{arcmin}^2$) and parallel NIRCam imaging through eight medium-band filters: F140M, F162M, F182M, F210M, F250M, F300M, F360M, and F460M (exposure times of 18--36 minutes per filter, covering $234\ \mathrm{arcmin}^2$; \citealt{Muzzin2025}). MINERVA reaches a $5\sigma$ depth of $\sim 28.1$ AB mag in F300M and $\sim 23.9$ AB mag in MIRI F1280W \citep{Muzzin2025}. Combining the core PRIMER and MINERVA observations in UDS yields 16 NIRCam filters (7 broadbands + 9 medium bands) and 4 MIRI filters (F770W, F1280W, F1500W, and F1800W), which together with 6 primary HST filters from the Cosmic Assembly Near-infrared Deep Extragalactic Legacy Survey (CANDELS) and 3D-HST (F435W, F606W, and F814W from the Advanced Camera for Surveys [ACS] and F125W, F140W, and F160W from the infrared channel of the Wide Field Camera 3 [WFC3/IR]) provide a unified baseline of 26 filters \citep{Muzzin2025}. Incorporating medium-band photometry substantially reduces the photometric redshift scatter (normalized median absolute deviation, $\sigma_{\mathrm{NMAD}}$) and catastrophic outlier rates compared to broadband-only measurements \citep{Suess2024, Muzzin2025}. 

\subsection{MINERVA SUPER Catalog (DR0)}

We utilize the value-added MINERVA SUPER Catalog from DR0\footnote{\url{https://minerva.colorado.edu/data-release.html}}, which provides multi-wavelength photometry across the UDS field \citep{Muzzin2025}. The MINERVA team constructed this catalog using a detection image generated from a noise-weighted stack of long-wavelength NIRCam filters (F277W, F356W, and F444W). The team reduced the NIRCam and HST mosaics with \texttt{grizli} \citep{Grizli2019, Grizli2021} and drizzled them to a uniform $0\farcs04\ \mathrm{pixel}^{-1}$ scale, whereas they reduced the MIRI mosaics with a custom version of the pipeline developed for the Systematic Mid-infrared Instrument Legacy Extragalactic Survey \citep[SMILES;][]{Alberts2024} and drizzled them to $0\farcs08\ \mathrm{pixel}^{-1}$. In addition to the core PRIMER and MINERVA datasets, the DR0 photometric catalog integrates archival and parallel imaging from HST (ACS: F435W, F606W, and F814W; WFC3/IR: F125W, F140W, and F160W; \citealt{Grogin2011, Koekemoer2011, Brammer2012, Skelton2014, Momcheva2016}).

To ensure accurate SED modeling, the MINERVA team matched the HST and NIRCam images to the F444W PSF and derived total MIRI fluxes from MIRI-to-F444W flux ratios measured in fixed reference apertures on the PSF-matched F444W image\footnote{\url{https://minerva.colorado.edu/data_release_products/uds/catalogs/MINERVA-UDSDR0_README.txt}}. We select objects using the \texttt{use\_phot == 1} and \texttt{use\_phot\_miri == TRUE} (where applicable) quality flags to ensure reliable photometry free from artifacts.

\subsection{DAWN JWST Archive}

The DAWN JWST Archive\footnote{\url{https://dawn-cph.github.io/dja/index.html}} (DJA) provides public JWST galaxy data that the DJA team processed using the \texttt{grizli} \citep{Grizli2019, Grizli2021} and \texttt{msaexp} \citep{brammer_msaexp} pipelines. We retrieve galaxies with grade 3 Near-Infrared Spectrograph (NIRSpec) data from the DJA catalog \citep{brammer_msaexp,Heintz2024,deGraaff2025RUBIES}. A grade 3 spectrum indicates a robust redshift from one or more emission or absorption features. Out of the 42,462 grade 3 spectra available (as of 2026 August 31), 7 are common with our sample. For these galaxies, we use the spectroscopic redshifts for SED modeling, as described in Section~\ref{sec:methods}.

\section{Methods} \label{sec:methods}
\subsection{Stellar Population Synthesis with \texttt{BAGPIPES}}

To determine the global physical properties of galaxies, such as stellar mass, SFR, and photometric redshift, we use SED modeling. By comparing observed photometry across multiple filters with synthetic spectra produced by stellar population synthesis models, we derive joint posterior distributions for these critical parameters, which form the basis of our quiescent galaxy sample selection.

We perform SED modeling using data from the MINERVA DR0 photometric catalog, incorporating the JWST/NIRCam broadband and medium-band filters, JWST/MIRI filters, as well as archival HST/ACS and HST/WFC3 filters where available, using \bagpipes \citep{Carnall2018}. \bagpipes provides a flexible framework for generating synthetic galaxy spectra from user-specified parameters and fitting those models to observational data. The framework incorporates BC03 Stellar Population Synthesis (SPS) models from the 2016 update of \citet{Bruzual2003}, combined with the \citet{Kroupa2001} initial mass function. We adopt the delayed tau model for the star formation history (SFH) and the Nov 2022 JWST filter curves from \texttt{sedpy} \citep{johnson_2021_4582723}. \bagpipes employs the \texttt{PyMultiNest} nested sampling algorithm \citep{Feroz2008, Feroz2009, Feroz2019, Buchner2014} to efficiently sample the parameter space and generate model SEDs. By extensively sampling the posterior distribution, \bagpipes simultaneously constrains the photometric redshift and SPS parameters, returning joint posterior estimates of redshift, stellar mass, star formation rate, gas-phase metallicity, and dust attenuation. We average the SFR, and hence the sSFR, over the most recent 10 Myr of the SFH instead of the \bagpipes default of 100 Myr.

For each galaxy, we mask filters with a signal-to-noise ratio (SNR) of $\mathrm{SNR} \leqslant 3$. For galaxies with available spectroscopic redshifts from the DJA, we fix their redshifts to the spectroscopic values during the fitting procedure. Table~\ref{tab:bagpipesparams} summarizes the parameters and prior distributions adopted for the \bagpipes modeling.

\begin{deluxetable}{lcl}[ht!]
  \tabletypesize{\scriptsize}
  \tablecaption{BAGPIPES fitting parameters\label{tab:bagpipesparams}}
  \tablehead{
    \colhead{Parameter} & \colhead{Value / Range} & \colhead{Description}
  }
  \startdata
  \cutinhead{Star Formation History (Delayed Tau Model)}
  age & [0.1, 15] & Time since star formation began in Gyr \\
  tau & [0.3, 10] & e-folding timescale of SFH decay in Gyr \\
  massformed & [4, 15]  & Log of total stellar mass formed wrt solar mass \\
  metallicity & [0.1, 2.5] & Log of stellar metallicity wrt solar metallicity\\
  \cutinhead{Dust Attenuation}
  type & ``Calzetti'' & Dust attenuation law \\
  eta & 2 & Multiplicative factor for the Calzetti law \\
  $A_V$ & [0, 6] & V-band attenuation in mag \\
  \cutinhead{Nebular Emission}
  logU & [$-4$, $-2$] & Log of the ionization parameter \\
  \cutinhead{Fit Instructions}
  redshift & [0, 15] & Redshift range for the fit \\
  \enddata
\end{deluxetable}

\subsection{Sample Selection Using BAGPIPES Results}

Having derived the physical properties of galaxies in the MINERVA DR0 catalog using \bgcomma, we select massive galaxies at cosmic noon that have ceased significant star formation according to the following criteria:

\begin{itemize}
  \item Stellar Mass: We select galaxies with stellar masses (50th percentile) $M_* > 10^{10}$ M$_\odot$ to focus on the most massive systems.
  \item Specific Star Formation Rate (sSFR): We classify galaxies with
        \begin{equation}
          \mathrm{sSFR} < \frac{0.2}{t_{\mathrm{age}}},
          \label{eq:ssfr_cut}
        \end{equation}
        where $t_{\mathrm{age}}$ is the age of the universe at the galaxy's redshift, as quiescent, indicating minimal ongoing star formation \citep{Franx2008, Gallazzi2014, Schreiber2018, Carnall2023a, Baker2025, Prasal2026}.
  \item Redshift Range: We restrict our sample to galaxies within the redshift range $2 < z < 3$ to target the epoch of interest for cosmic noon studies.
  \item Number of High-SNR Photometric Bands: We require that galaxies have at least 14 photometric bands with $\mathrm{SNR} > 3$ to ensure robust SED modeling.
\end{itemize}

\begin{figure*}[ht!]
  \centering
  \includegraphics[width=\textwidth, height=0.85\textheight, keepaspectratio]{ 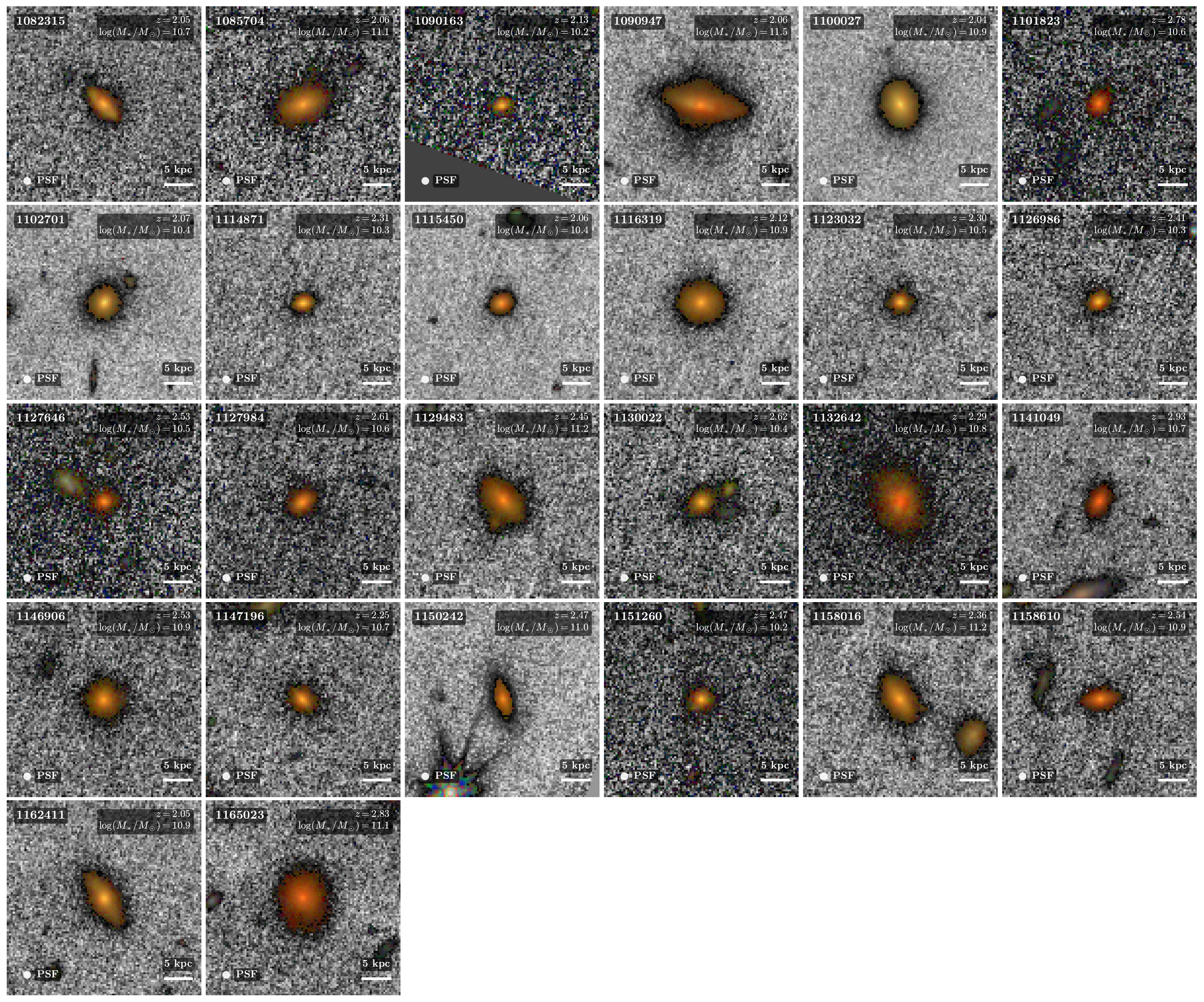}
  \caption{RGB images ($4''\times 4''$) of our 26 massive quiescent galaxies at $2 < z < 3$ in the MINERVA UDS field, made with NIRCam filters F200W (red), F150W (green), and F115W (blue) and centered on the galaxies. The white circle at the bottom left represents the PSF size of the F444W filter image (full width at half maximum, FWHM $= 0\farcs157$).}
  \label{fig:galaxy_mosaic}
\end{figure*}

\begin{figure}[ht!]
  \centering
  \includegraphics[width=0.85\columnwidth, height=0.6\textheight, keepaspectratio]{ 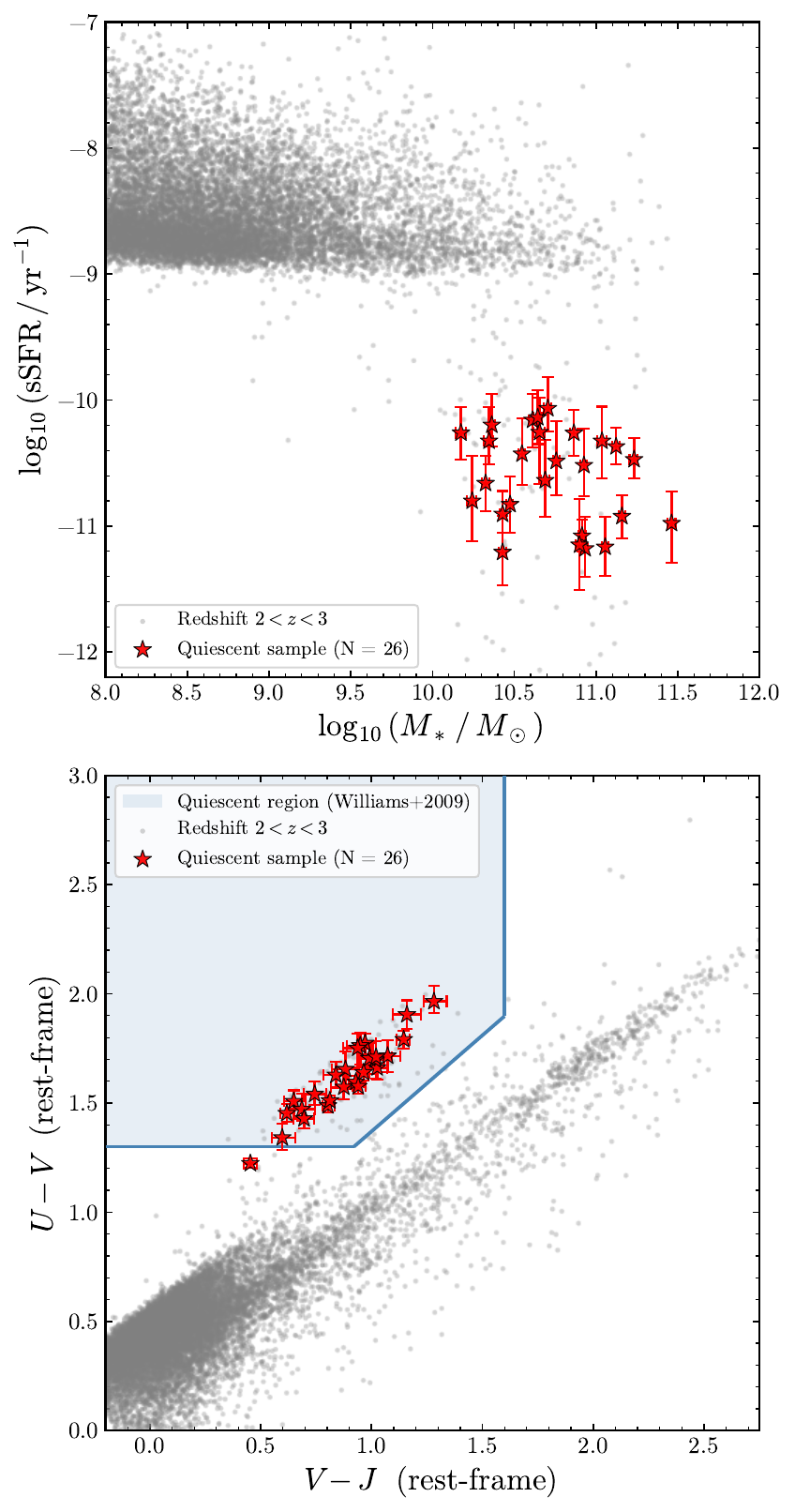}
  \caption{Stellar mass vs.\ sSFR (in log scale, top panel) and $UVJ$ diagram (bottom panel) for our sample of 26 massive quiescent galaxies and all survey galaxies at $2 < z < 3$. The quiescent selection boundary from \citet{Williams2009} is shown in the $UVJ$ plot as a light-blue shaded region, with our sample galaxies (red stars) lying well within the quiescent region. The stellar mass vs sSFR plot shows that our sample galaxies lie well below the star-forming galaxies, consistent with their quiescent nature. The gray points represent all galaxies in the redshift range $2 < z < 3$ in the MINERVA field.}
  \label{fig:sfms_uvj}
\end{figure}

Furthermore, we only use the galaxies which have medium band coverage in the MINERVA DR0 catalog. This is done to ensure the robustness of the SED modeling and to separate our study from the previous studies \citep{Haryana_2025,Chen2026}. Applying these selection criteria yields a final sample of 26 massive quiescent galaxies, displayed as RGB cutouts in Figure~\ref{fig:galaxy_mosaic}. Figure~\ref{fig:sfms_uvj} illustrates the distribution of our sample in the stellar mass vs.\ sSFR plane and the rest-frame $U-V$ vs.\ $V-J$ ($UVJ$) color diagram, demonstrating that our targets cleanly satisfy empirical quiescence boundaries (\id = 1130022 lies outside the quiescent region defined by \citet{Williams2009}, but we consider it in our sample since it satisfies the sSFR quiescence criterion). Table~\ref{tab:final_samples} presents the estimated physical and SPS parameters for all 26 galaxies in the final sample.

In the literature, two popular parameterizations describe galaxy Star Formation Histories: the delayed tau model \citep{Sato2024, Haryana_2025,Prasal2026} and the double power-law model \citep{Carnall2018, Carnall2024, Baker2025}. We adopt the delayed tau model for our analysis. When comparing the derived parameters with those from a double power-law SFH, we find no significant differences. To evaluate the robustness of our methodology, we also test a non-parametric SFH following \citet{Leja2019}. Since \bagpipes requires a fixed redshift to constrain non-parametric SFHs, we test the 7 galaxies with confirmed spectroscopic redshifts (\texttt{ID} = 1101823, 1127646, 1127984, 1130022, 1132642, 1141049, and 1146906) using the non-parametric SFH, confirming that all 7 robustly satisfy the quiescence criterion (Equation~\ref{eq:ssfr_cut}). Also, our derived size-mass relations remain within $1 \ \sigma$ uncertainties for the double power law SFH.

\begin{deluxetable*}{ccccccccccc}[ht!]
  \tabletypesize{\scriptsize}
  \tablecaption{Physical properties of final sample galaxies from BAGPIPES SED fitting\label{tab:final_samples}}
  \tablehead{
    \colhead{ID} & \colhead{Age} & \colhead{log($M_{\mathrm{formed}}/M_\odot$)} & \colhead{Metallicity} & \colhead{$\tau$} & \colhead{$A_V$} & \colhead{Redshift} & \colhead{log($M_*/M_\odot$)} & \colhead{SFR} & \colhead{log(sSFR/yr$^{-1}$)} & \colhead{Mass Weighted Age} \\
    \colhead{} & \colhead{(Gyr)} & \colhead{} & \colhead{($Z_\odot$)} & \colhead{(Gyr)} & \colhead{(mag)} & \colhead{} & \colhead{} & \colhead{($M_\odot\ \mathrm{yr}^{-1}$)} & \colhead{} & \colhead{(Gyr)}
  }
  \startdata
  1082315 & $2.46_{-0.45}^{+0.48}$ & $10.91_{-0.04}^{+0.04}$ & $0.486_{-0.223}^{+0.634}$ & $0.37_{-0.05}^{+0.10}$ & $0.66_{-0.41}^{+0.18}$ & $2.05_{-0.06}^{+0.05}$ & $10.65_{-0.03}^{+0.03}$ & $2.45_{-1.46}^{+2.20}$ & $-10.26_{-0.41}^{+0.27}$ & $1.71_{-0.35}^{+0.36}$ \\
  1085704 & $2.95_{-0.20}^{+0.15}$ & $11.33_{-0.02}^{+0.02}$ & $0.329_{-0.127}^{+0.295}$ & $0.32_{-0.02}^{+0.03}$ & $0.46_{-0.20}^{+0.16}$ & $2.06_{-0.05}^{+0.05}$ & $11.06_{-0.02}^{+0.02}$ & $0.78_{-0.33}^{+0.60}$ & $-11.17_{-0.23}^{+0.24}$ & $2.29_{-0.19}^{+0.14}$ \\
  1090163 & $2.74_{-0.33}^{+0.27}$ & $10.51_{-0.03}^{+0.03}$ & $0.315_{-0.117}^{+0.224}$ & $0.34_{-0.03}^{+0.04}$ & $0.40_{-0.18}^{+0.17}$ & $2.13_{-0.10}^{+0.11}$ & $10.24_{-0.03}^{+0.03}$ & $0.28_{-0.15}^{+0.34}$ & $-10.80_{-0.32}^{+0.36}$ & $2.05_{-0.31}^{+0.27}$ \\
  1090947 & $2.88_{-0.26}^{+0.22}$ & $11.74_{-0.03}^{+0.03}$ & $0.227_{-0.070}^{+0.158}$ & $0.33_{-0.03}^{+0.04}$ & $1.04_{-0.15}^{+0.15}$ & $2.06_{-0.08}^{+0.06}$ & $11.46_{-0.03}^{+0.02}$ & $3.07_{-1.57}^{+2.45}$ & $-10.98_{-0.31}^{+0.25}$ & $2.21_{-0.24}^{+0.18}$ \\
  1100027 & $2.88_{-0.18}^{+0.23}$ & $11.19_{-0.03}^{+0.03}$ & $0.218_{-0.040}^{+0.072}$ & $0.33_{-0.02}^{+0.03}$ & $0.09_{-0.06}^{+0.08}$ & $2.04_{-0.08}^{+0.05}$ & $10.91_{-0.02}^{+0.02}$ & $0.67_{-0.17}^{+0.25}$ & $-11.08_{-0.13}^{+0.13}$ & $2.23_{-0.15}^{+0.17}$ \\
  1101823\tablenotemark{a} & $2.08_{-0.18}^{+0.15}$ & $10.86_{-0.03}^{+0.02}$ & $0.301_{-0.100}^{+0.138}$ & $0.33_{-0.02}^{+0.03}$ & $0.70_{-0.15}^{+0.11}$ & 2.78 & $10.61_{-0.02}^{+0.02}$ & $2.64_{-0.89}^{+1.78}$ & $-10.16_{-0.22}^{+0.21}$ & $1.45_{-0.18}^{+0.13}$ \\
  1102701 & $2.95_{-0.18}^{+0.11}$ & $10.71_{-0.02}^{+0.02}$ & $0.181_{-0.040}^{+0.047}$ & $0.36_{-0.02}^{+0.03}$ & $0.11_{-0.07}^{+0.10}$ & $2.07_{-0.05}^{+0.05}$ & $10.43_{-0.02}^{+0.02}$ & $0.34_{-0.10}^{+0.18}$ & $-10.90_{-0.15}^{+0.18}$ & $2.24_{-0.16}^{+0.11}$ \\
  1114871 & $2.50_{-0.17}^{+0.21}$ & $10.59_{-0.03}^{+0.03}$ & $0.231_{-0.052}^{+0.109}$ & $0.33_{-0.02}^{+0.03}$ & $0.22_{-0.11}^{+0.10}$ & $2.31_{-0.09}^{+0.06}$ & $10.32_{-0.02}^{+0.03}$ & $0.46_{-0.18}^{+0.28}$ & $-10.66_{-0.22}^{+0.22}$ & $1.85_{-0.17}^{+0.18}$ \\
  1115450 & $2.98_{-0.19}^{+0.13}$ & $10.70_{-0.02}^{+0.03}$ & $0.790_{-0.460}^{+0.737}$ & $0.32_{-0.02}^{+0.03}$ & $0.44_{-0.29}^{+0.31}$ & $2.06_{-0.08}^{+0.07}$ & $10.43_{-0.02}^{+0.02}$ & $0.17_{-0.08}^{+0.16}$ & $-11.21_{-0.26}^{+0.29}$ & $2.31_{-0.19}^{+0.15}$ \\
  1116319 & $2.89_{-0.20}^{+0.16}$ & $11.21_{-0.02}^{+0.02}$ & $0.419_{-0.198}^{+0.306}$ & $0.32_{-0.01}^{+0.03}$ & $0.37_{-0.17}^{+0.14}$ & $2.12_{-0.06}^{+0.09}$ & $10.93_{-0.02}^{+0.02}$ & $0.56_{-0.23}^{+0.49}$ & $-11.18_{-0.23}^{+0.25}$ & $2.25_{-0.19}^{+0.15}$ \\
  1123032 & $2.59_{-0.20}^{+0.17}$ & $10.75_{-0.03}^{+0.02}$ & $0.214_{-0.037}^{+0.061}$ & $0.32_{-0.01}^{+0.02}$ & $0.29_{-0.10}^{+0.11}$ & $2.30_{-0.09}^{+0.05}$ & $10.47_{-0.02}^{+0.02}$ & $0.44_{-0.18}^{+0.29}$ & $-10.83_{-0.22}^{+0.22}$ & $1.96_{-0.19}^{+0.16}$ \\
  1126986 & $2.26_{-0.27}^{+0.28}$ & $10.61_{-0.04}^{+0.03}$ & $0.169_{-0.040}^{+0.041}$ & $0.34_{-0.03}^{+0.04}$ & $0.41_{-0.13}^{+0.16}$ & $2.41_{-0.07}^{+0.06}$ & $10.35_{-0.03}^{+0.03}$ & $1.04_{-0.34}^{+0.84}$ & $-10.32_{-0.19}^{+0.27}$ & $1.61_{-0.24}^{+0.22}$ \\
  1127646\tablenotemark{a} & $2.36_{-0.20}^{+0.11}$ & $10.80_{-0.02}^{+0.02}$ & $0.540_{-0.259}^{+0.643}$ & $0.33_{-0.02}^{+0.05}$ & $0.50_{-0.32}^{+0.21}$ & 2.53 & $10.55_{-0.02}^{+0.02}$ & $1.31_{-0.56}^{+1.19}$ & $-10.43_{-0.25}^{+0.28}$ & $1.69_{-0.17}^{+0.12}$ \\
  1127984\tablenotemark{a} & $2.07_{-0.19}^{+0.24}$ & $10.89_{-0.03}^{+0.04}$ & $0.244_{-0.079}^{+0.123}$ & $0.33_{-0.02}^{+0.05}$ & $0.53_{-0.12}^{+0.18}$ & 2.61 & $10.64_{-0.03}^{+0.02}$ & $3.10_{-1.08}^{+2.09}$ & $-10.14_{-0.21}^{+0.22}$ & $1.45_{-0.20}^{+0.16}$ \\
  1129483 & $2.33_{-0.17}^{+0.17}$ & $11.50_{-0.03}^{+0.02}$ & $0.172_{-0.029}^{+0.033}$ & $0.33_{-0.02}^{+0.03}$ & $0.85_{-0.08}^{+0.10}$ & $2.45_{-0.03}^{+0.03}$ & $11.23_{-0.03}^{+0.02}$ & $5.75_{-1.60}^{+2.71}$ & $-10.47_{-0.15}^{+0.17}$ & $1.68_{-0.13}^{+0.15}$ \\
  1130022\tablenotemark{a} & $2.01_{-0.21}^{+0.19}$ & $10.62_{-0.03}^{+0.02}$ & $0.130_{-0.020}^{+0.031}$ & $0.31_{-0.01}^{+0.02}$ & $0.19_{-0.12}^{+0.15}$ & 2.62 & $10.36_{-0.02}^{+0.02}$ & $1.48_{-0.49}^{+0.98}$ & $-10.20_{-0.17}^{+0.25}$ & $1.40_{-0.20}^{+0.17}$ \\
  1132642\tablenotemark{a} & $2.59_{-0.25}^{+0.16}$ & $11.02_{-0.03}^{+0.03}$ & $0.479_{-0.260}^{+0.428}$ & $0.36_{-0.04}^{+0.07}$ & $0.77_{-0.24}^{+0.23}$ & 2.29 & $10.76_{-0.02}^{+0.02}$ & $1.92_{-0.91}^{+1.88}$ & $-10.48_{-0.27}^{+0.32}$ & $1.84_{-0.20}^{+0.15}$ \\
  1141049\tablenotemark{a} & $1.99_{-0.20}^{+0.13}$ & $10.95_{-0.03}^{+0.03}$ & $0.207_{-0.067}^{+0.107}$ & $0.33_{-0.02}^{+0.04}$ & $0.80_{-0.13}^{+0.20}$ & 2.92 & $10.71_{-0.02}^{+0.02}$ & $4.33_{-1.43}^{+3.23}$ & $-10.07_{-0.18}^{+0.25}$ & $1.35_{-0.18}^{+0.12}$ \\
  1146906\tablenotemark{a} & $2.16_{-0.19}^{+0.19}$ & $11.12_{-0.03}^{+0.03}$ & $0.174_{-0.047}^{+0.059}$ & $0.33_{-0.02}^{+0.03}$ & $0.60_{-0.13}^{+0.15}$ & 2.53 & $10.87_{-0.03}^{+0.02}$ & $3.93_{-1.27}^{+2.34}$ & $-10.26_{-0.18}^{+0.19}$ & $1.52_{-0.17}^{+0.16}$ \\
  1147196 & $2.62_{-0.27}^{+0.19}$ & $10.95_{-0.03}^{+0.03}$ & $0.342_{-0.141}^{+0.431}$ & $0.34_{-0.03}^{+0.05}$ & $0.60_{-0.29}^{+0.15}$ & $2.25_{-0.08}^{+0.08}$ & $10.69_{-0.03}^{+0.02}$ & $1.12_{-0.53}^{+1.23}$ & $-10.64_{-0.29}^{+0.32}$ & $1.92_{-0.23}^{+0.18}$ \\
  1150242 & $2.26_{-0.24}^{+0.24}$ & $11.29_{-0.03}^{+0.04}$ & $0.453_{-0.205}^{+0.648}$ & $0.33_{-0.02}^{+0.04}$ & $0.73_{-0.27}^{+0.16}$ & $2.47_{-0.07}^{+0.07}$ & $11.04_{-0.03}^{+0.03}$ & $5.23_{-2.57}^{+4.33}$ & $-10.32_{-0.29}^{+0.27}$ & $1.59_{-0.20}^{+0.23}$ \\
  1151260 & $2.19_{-0.24}^{+0.23}$ & $10.42_{-0.04}^{+0.04}$ & $0.276_{-0.086}^{+0.133}$ & $0.33_{-0.02}^{+0.05}$ & $0.17_{-0.11}^{+0.14}$ & $2.47_{-0.07}^{+0.08}$ & $10.17_{-0.03}^{+0.03}$ & $0.81_{-0.29}^{+0.45}$ & $-10.26_{-0.21}^{+0.21}$ & $1.55_{-0.20}^{+0.18}$ \\
  1158016 & $2.63_{-0.12}^{+0.12}$ & $11.43_{-0.02}^{+0.02}$ & $0.229_{-0.040}^{+0.115}$ & $0.31_{-0.01}^{+0.02}$ & $0.38_{-0.10}^{+0.09}$ & $2.36_{-0.07}^{+0.08}$ & $11.16_{-0.02}^{+0.02}$ & $1.70_{-0.55}^{+0.85}$ & $-10.92_{-0.18}^{+0.17}$ & $2.00_{-0.11}^{+0.13}$ \\
  1158610 & $2.38_{-0.22}^{+0.14}$ & $11.18_{-0.03}^{+0.02}$ & $0.428_{-0.211}^{+0.520}$ & $0.32_{-0.02}^{+0.03}$ & $0.62_{-0.26}^{+0.17}$ & $2.54_{-0.08}^{+0.12}$ & $10.93_{-0.02}^{+0.02}$ & $2.52_{-1.06}^{+2.38}$ & $-10.52_{-0.24}^{+0.29}$ & $1.73_{-0.21}^{+0.14}$ \\
  1162411 & $2.95_{-0.32}^{+0.20}$ & $11.17_{-0.03}^{+0.03}$ & $0.341_{-0.129}^{+0.234}$ & $0.32_{-0.02}^{+0.03}$ & $0.39_{-0.16}^{+0.13}$ & $2.05_{-0.09}^{+0.09}$ & $10.90_{-0.02}^{+0.02}$ & $0.55_{-0.30}^{+0.77}$ & $-11.15_{-0.36}^{+0.36}$ & $2.28_{-0.30}^{+0.21}$ \\
  1165023 & $2.16_{-0.14}^{+0.10}$ & $11.38_{-0.03}^{+0.02}$ & $0.214_{-0.055}^{+0.084}$ & $0.31_{-0.01}^{+0.01}$ & $0.60_{-0.08}^{+0.10}$ & $2.83_{-0.09}^{+0.08}$ & $11.12_{-0.02}^{+0.02}$ & $5.70_{-1.55}^{+2.37}$ & $-10.37_{-0.14}^{+0.15}$ & $1.54_{-0.13}^{+0.10}$ \\
  \enddata
  \tablecomments{Values shown as median with 16th/84th percentile uncertainties $\mathrm{value}_{-\sigma_{\mathrm{lo}}}^{+\sigma_{\mathrm{hi}}}$. All physical parameters have been derived from BAGPIPES SED fitting with delayed tau star formation history.}
  \tablenotetext{a}{Galaxies with confirmed spectroscopic redshifts from the DJA. We do not show redshift uncertainties for these objects.}
\end{deluxetable*}

\subsection{Empirical PSF Construction and Convolution Kernel Generation} \label{sec:psf_homogenization}

To ensure consistent spatial resolution across all bands, we construct high signal-to-noise empirical PSFs directly from the MINERVA science mosaics and generate regularized Fourier convolution kernels to match all shorter-wavelength filters to the F444W filter, adopting the methodology of \citet{Weaver2024} adapted from \citet{Skelton2014} and \citet{Whitaker2019}.

We build empirical PSFs from isolated, unsaturated stars within the science mosaics rather than using simulated PSFs from \texttt{WebbPSF} \citep{Perrin2012, Perrin2014}. While simulated PSFs capture nominal telescope optics, \citet{Weaver2024} demonstrated that \texttt{WebbPSF} models are $1\%$--$2\%$ narrower than real stars observed in drizzled mosaics (see Appendix~\ref{sec:appendix_psf} for a comprehensive forward-drizzling comparison across all 15 filters).

We construct empirical PSFs for each NIRCam filter through an automated, multi-step pipeline. Firstly, we identify candidate point sources in the MINERVA DR0 catalog using the stellar classification flag (\texttt{flag\_star == 1}). To avoid edge artifacts, we reject candidates located within 150 pixels of the mosaic boundary or exhibiting zero weight ($w = 0$) within a $100$-pixel ($4\farcs0$) radius. We inspect the central $15 \times 15$ pixel core ($r \leqslant 7$ pixels) of each candidate and discard saturated stars containing non-finite pixels or $w \leqslant 0$. Secondly, we visually inspect the accepted candidates and remove any galaxies that were misclassified as stars in the catalog. For each accepted candidate, we extract a $301 \times 301$ pixel cutout to estimate the local sky background in source-free pixels ($\texttt{seg} == 0$) located at $r > 100$ pixels ($4\farcs0$) from the star center, bypassing the extended Airy rings. If the background shows a significant spatial gradient across four quadrants (median spread $> 0.5\sigma_{\mathrm{bg}}$), we subtract a two-dimensional background model using \texttt{Background2D} from \texttt{photutils} \citep{Bradley2025}. After that, we mask neighboring segmented objects and discard candidates if the contaminant flux has $\mathrm{SNR} \geqslant 5.0$ and exceeds $3\%$ of the target stellar flux within the central $101 \times 101$ pixel stamp.

Following the size--magnitude locus selection method \citep{Skelton2014, Whitaker2019, Weaver2024}, we isolate true point sources in the size--magnitude plane using the $0\farcs32$ aperture magnitude ($m_{0.32}$) and the concentration flux ratio $R_{16/32} = F(0\farcs16) / F(0\farcs32)$. We first reject candidates fainter than $m_{0.32} = 24.0$ AB mag, then fit a linear relation to the stellar locus of the remaining stars and reject those deviating by $> 2.0\ \sigma$ from it, a stricter cut than the $2.8\ \sigma$ adopted by \citet{Weaver2024}. From the surviving candidates, we select up to 60 brightest stars for broadband filters and 30 for medium-band filters, and extract a $201 \times 201$ pixel stamp around each. We determine the subpixel centroid of each star within a central $21 \times 21$ pixel window using a center-of-mass algorithm, translate the star to a common grid center via cubic spline interpolation, and normalize each stamp to unit enclosed flux within a 16-pixel ($0\farcs64$) diameter circular aperture \citep{Weaver2024}. Pixels belonging to neighboring sources are masked as \texttt{NaN} after interpolation to prevent flux contamination. Finally, we combine the aligned stamps using a $3.0\ \sigma$-clipped mean stack, renormalize the stacked PSF within a $4\farcs0$ diameter aperture such that the energy enclosed within a 4\arcsec diameter aligns with typical calibration levels\footnote{\url{https://jwst-docs.stsci.edu/jwst-near-infrared-camera/nircam-performance/nircam-point-spread-functions}}, and crop the result to a final $101 \times 101$ pixel stamp ($4\farcs04 \times 4\farcs04$ at $0\farcs04\ \mathrm{pixel}^{-1}$).

To homogenize image resolution across all bands, we generate convolution kernels matching each shorter-wavelength empirical PSF ($P_{\mathrm{src}}$) to the F444W empirical PSF ($P_{\mathrm{tgt}}$) using \texttt{PyPHER} \citep{Boucaud2016}. \texttt{PyPHER} computes the matching kernel $\widetilde{K}(\mathbf{k})$ via regularized Wiener filter deconvolution \citep{Wiener1949, Boucaud2016} in Fourier space:
\begin{equation}
  \widetilde{K}(\mathbf{k}) = \frac{\widetilde{P}_{\mathrm{tgt}}(\mathbf{k}) \cdot \widetilde{P}_{\mathrm{src}}^*(\mathbf{k})}{|\widetilde{P}_{\mathrm{src}}(\mathbf{k})|^2 + r \cdot |\widetilde{D}(\mathbf{k})|^2},
  \label{eq:pypher_kernel}
\end{equation}
where $\widetilde{P}^*$ represents the complex conjugate, $\widetilde{D}(\mathbf{k})$ is the Fourier transform of the 2D discrete Laplacian operator ($D$) that penalizes high-frequency noise \citep{Boucaud2016}, and $r$ is a dimensionless regularization parameter. Following \citet{Weaver2024}, we perform the kernel matching in $3\times$ oversampled space ($0\farcs0133\ \mathrm{pixel}^{-1}$), after which we rescale the kernels to the original pixel scale ($0\farcs04\ \mathrm{pixel}^{-1}$). We adopt a regularization parameter of $r = 3 \times 10^{-3}$, which avoids high-frequency noise in the kernels while maintaining deviations below $1\%$ across all aperture diameters of interest \citep{Weaver2024}. We normalize the resulting kernels to unit sum ($\sum_{x,y} K(x,y) = 1.0$) to guarantee flux conservation.

\begin{figure}[ht!]
  \centering
  \includegraphics[width=\columnwidth]{ 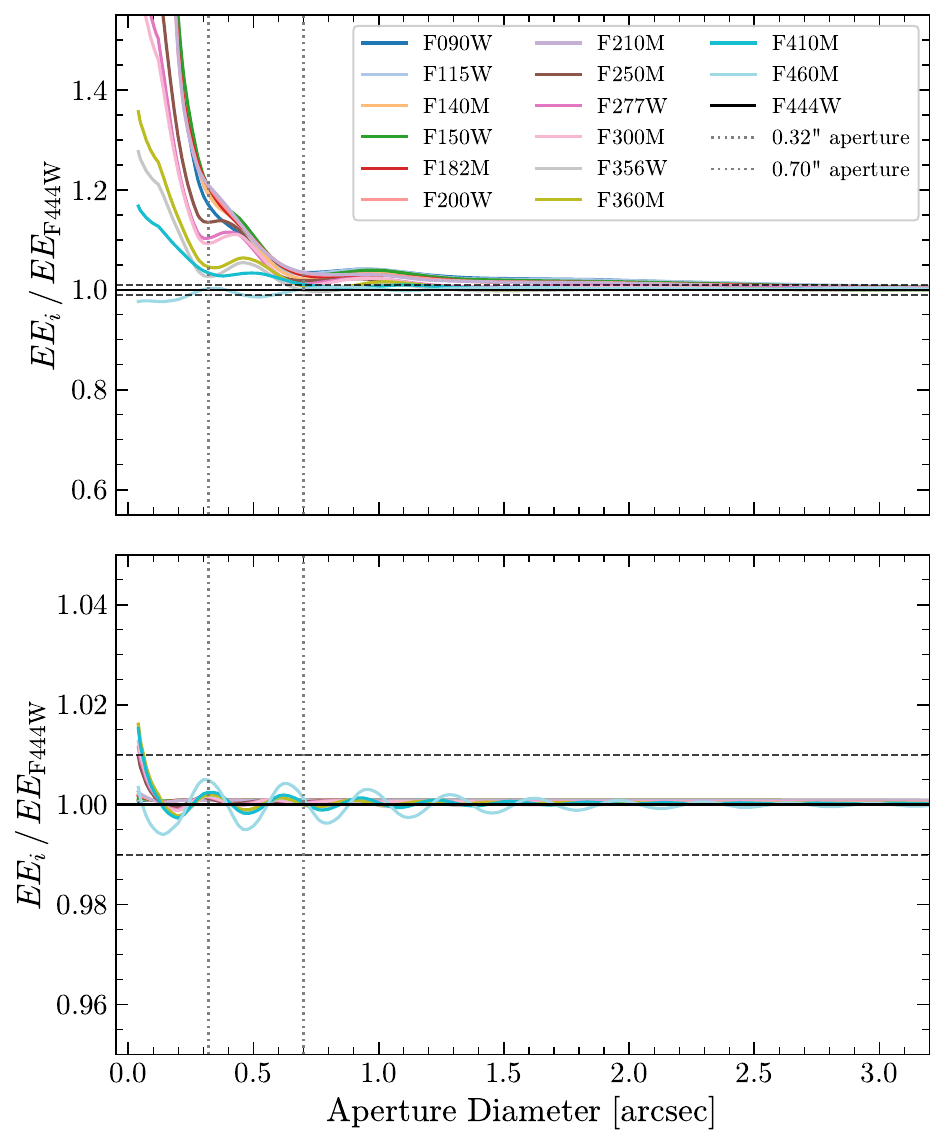}
  \caption{Encircled energy growth curve ratios $EE_i/ EE_{\mathrm{F444W}}$ as a function of circular aperture diameter for all 14 NIRCam filters matched to the target F444W empirical PSF, following \citet[][Figure~6]{Weaver2024}. Top panel: Pre-convolution flux ratios, showing pronounced central light concentration in shorter-wavelength filters ($EE_i/EE_{\mathrm{F444W}} > 1.2$). Bottom panel: Post-convolution flux ratios after applying \texttt{PyPHER} regularized Fourier matching kernels. All curves lie tightly within $\pm 1\%$ of unity (horizontal dashed lines) across all radii, including the fiducial aperture diameters at $0\farcs32$ and $0\farcs70$ (vertical dotted lines).}
  \label{fig:combined_growth_curves}
\end{figure}

We validate kernel performance by convolving each empirical PSF with its matching kernel via fast Fourier transforms and comparing the resulting encircled energy (EE) growth curves to the target F444W PSF (Figure~\ref{fig:combined_growth_curves}). Across all filters, the post-convolution encircled energy ratios align tightly with unity, with residual deviations $|EE_{\mathrm{filter}}(r) / EE_{\mathrm{F444W}}(r) - 1.0| < 0.01$ ($< 1\%$) across both standard photometric apertures ($0\farcs32$ and $0\farcs70$ diameters).

\subsection{Morphological Modeling and Size--Mass Scaling Methodology} \label{sec:sizemass_methods}

To characterize the spatial distribution of stellar light, structural properties, and size--mass scaling relations of our sample of massive quiescent galaxies, we perform parametric morphological modeling across six JWST/NIRCam broadband filters: F115W, F150W, F200W, F277W, F356W, and F444W. At our median sample redshift $z \approx 2.33$, the F200W filter probes rest-frame optical wavelengths ($\lambda_{\mathrm{rest}} \approx 0.60\ \mu\mathrm{m}$, corresponding to rest-frame $V/r$-band) and the F356W filter probes rest-frame near-infrared wavelengths ($\lambda_{\mathrm{rest}} \approx 1.10\ \mu\mathrm{m}$, corresponding to rest-frame $Y/J$-band). This wavelength coverage enables direct structural comparisons with various high-redshift studies \citep[e.g.,][]{vanderWel2014, Mowla2019b, Kawinwanichakij2026, Ito2024}. Although we fit all six filters (Appendix~\ref{sec:appendix_morphology}), we focus the detailed analysis in the following sections on F200W, which traces the rest-frame optical light, and F356W, which traces the rest-frame near-infrared light and provides a closer proxy for the underlying stellar mass distribution. This approach follows previous studies that compare galaxy structure between one rest-frame optical and one rest-frame near-infrared band \citep[e.g.,][]{Cutler2024, Kawinwanichakij2026}.

\subsubsection{Parametric Morphological Modeling with \texttt{pysersic}} \label{sec:pysersic_fitting}

For parametric morphological analysis, we use the \pysersic python package \citep{Pasha2023} to model the light profiles and perform structural decompositions of the galaxies. This Python library fits \sersic (and other) profiles to astronomical images using Bayesian inference, accelerated via JAX.

Firstly, we prepare $6\farcs0 \times 6\farcs0$ ($150 \times 150$ pixel) cutouts of the broadband filter images (F115W, F150W, F200W, F277W, F356W, and F444W) centered on each galaxy from the science and weight mosaics, together with their filter-matched empirical PSFs (Section~\ref{sec:psf_homogenization}). In our \pysersic pipeline, we explicitly mask out pixels with invalid values in the weight maps (where weight $\leqslant 0$ or non-finite). These pixels are added to the mask to exclude them from the fit. As an additional safeguard, we set the RMS values for these invalid pixels to $10\times$ the median RMS of the valid pixels, ensuring they do not bias the model. Furthermore, to prevent contamination from neighboring objects, we perform multi-source detection using the python library for Source Extraction and Photometry (\texttt{SEP}; \citealt{Bertin1996, Barbary2016}) with a detection threshold of $3.0\ \sigma$ and a deblending contrast ratio of $\texttt{deblend\_cont} = 0.001$. Rather than masking adjacent sources, we simultaneously model all detected nearby sources as \sersic profiles alongside the primary target galaxy using \texttt{FitMulti}, explicitly fitting their overlapping light profiles. We note that galaxy \id = 1150242 has a star near it. So, we fit a \texttt{pointsource} profile for the star. On the other hand, \id = 1090947 has an uneven and asymmetric morphology, which is difficult to model with a light profile.

We then fit two structural configurations in \texttt{pysersic}: a single \sersic model \citep{Sersic1968} and a two-component \texttt{sersic\_exp} model. The \texttt{sersic\_exp} model is a combination of a \sersic profile \citep[for the bulge]{Sersic1968} and an exponential profile \citep[for the disk with $n_{\mathrm{disk}} = 1$;][]{Freeman1970}. From the \texttt{sersic\_exp} fit, we compute the bulge-to-total flux ratio ($B/T$), classifying galaxies as bulge-dominated if $B/T \geqslant 0.5$ and disk-dominated if $B/T < 0.5$. We construct custom multi-component priors via \texttt{PySersicMultiPrior} using the source positions, fluxes, sizes, and orientations measured by \texttt{SEP}. To prevent centroiding errors and degeneracies with close companions, we constrain the $X_c$ and $Y_c$ centroid priors for the target galaxy with uniform distributions within a $\pm 5.0$ pixel ($\pm 0\farcs20$) window centered on the \texttt{SEP}-detected coordinates. We adopt the Student-$t$ loss function to improve robustness against non-Gaussian noise outliers and residual substructure. We use \texttt{sky\_type} =``flat'' and estimate the posterior distribution using the svi-flow method, a form of variational inference that uses the block neural autoregressive flow (BNAF), a normalizing flow model, to approximate the posterior distribution \citep{decao2019blockneuralautoregressiveflow}.

\subsubsection{Size--Mass Relation Modeling} \label{sec:sizemass_fitting}

For modeling the size--mass relation analytically, we follow the methodology of \citet{Kawinwanichakij2026} \citep[see also][]{Shen2003,vanderWel2014, Mowla2019b, Kawinwanichakij2021, Nedkova2021, Ito2024}. We model the intrinsic distribution of galaxy sizes by assuming a log-normal form, parameterized as $\mathcal{N}(\log_{10} r, \sigma_{\log r})$, where $r$ represents the mean intrinsic effective radius and $\sigma_{\log r}$ is the intrinsic scatter of the distribution. This is different from the measured galaxy size via the half-light radius, $R_e$, which is affected by measurement uncertainties, $\delta \log_{10} R_e$. We parameterize the mean intrinsic size $r$ as a function of stellar mass as:
\begin{equation}
  \frac{r}{\mathrm{kpc}} = A \left( \frac{M_*}{5 \times 10^{10}\ M_\odot} \right)^\alpha,
  \label{eq:size_mass_model}
\end{equation}
where $A$ is the effective radius at $M_* = 5 \times 10^{10}\ M_\odot$ and $\alpha$ is the slope of the size--mass relation. Additionally, we assume that $\sigma_{\log r}$ is independent of mass.

Thereafter, we compute the total likelihood of the three parameters ($A, \alpha$, and $\sigma_{\log r}$) as:
\begin{equation}
  \begin{aligned}
    \ln \mathcal{L} = \sum_{i=1}^N \ln \big[ & W(M_{*,i}) \cdot \langle \mathcal{N}(\log_{10} r_i, \sigma_{\log r}) \cdot \\
                                         & \mathcal{N}(\log_{10} R_{e,i}, \delta \log R_{e,i}) \rangle \big],
  \end{aligned}
  \label{eq:total_sample_likelihood}
\end{equation}
where $W(M_*)$ is a weighting factor implemented to prevent the fit from being dominated by the more numerous low-mass galaxies \citep{vanderWel2014}. We compute $W(M_*)$ as the inverse of the stellar mass function (SMF) of quiescent galaxies at $2.0 < z \leqslant 2.5$, following \citet{Weaver2023}. The SMF is parameterized by a Schechter function \citep{Schechter1976}:
\begin{equation}
  \Phi(M_*) \propto \left( \frac{M_*}{M_0} \right)^{\alpha_{\mathrm{SMF}}+1} \exp\left( -\frac{M_*}{M_0} \right),
  \label{eq:schechter_smf}
\end{equation}
where we adopt $\log_{10}(M_0/M_\odot) = 10.49$ and $\alpha_{\mathrm{SMF}} = -0.66$ from \citet[Table~C.3;][]{Weaver2023}. The weight is defined as:
\begin{equation}
  W(M_*) = \frac{1}{\Phi(M_*)},
\end{equation}

This ensures that all stellar mass bins contribute appropriately to the likelihood calculation.

Several studies at lower redshifts report that the size--mass relation flattens below a pivot mass of $\log_{10}(M_*/M_\odot) \sim 10.3$ \citep[e.g.,][]{Mowla2019a, Kawinwanichakij2021, Nedkova2021, Cutler2022}, implying that a single power-law form (Equation~\ref{eq:size_mass_model}) may not be adequate to describe the full stellar mass range. To circumvent this, we fit the size--mass relation using only galaxies with $\log_{10}(M_*/M_\odot) > 10.3$ ($N = 24$ galaxies in our sample), excluding the two lowest-mass systems (ID~1151260 with $\log M_* = 10.17$ and ID~1090163 with $\log M_* = 10.24$).\footnote{Including these two galaxies changes $\alpha$ by $\leq 0.02$ and leaves $\log A$ and $\sigma_{\log r}$ essentially unchanged in all six filters.}

We fit the observed $R_e - M_*$ relation and estimate posterior distributions of the parameters ($A, \alpha, \sigma_{\log r}$) using \texttt{dynesty} nested sampling \citep{Skilling2004,Skilling2006,Speagle2020,Koposov2026}. We assume flat priors on all parameters: $\alpha \in [0, 1]$, $\log_{10}(A/\mathrm{kpc}) \in [-2, 1]$, and $\sigma_{\log r} \in [0.05, 0.50]\ \mathrm{dex}$. Table~\ref{tab:sizemass_params} summarizes the best-fit parameters across all six JWST/NIRCam filters.

\begin{deluxetable*}{lccccc}[ht!]
  \tabletypesize{\scriptsize}
  \tablecaption{Best-fit Size--Mass Relation Parameters across JWST NIRCam Filters\label{tab:sizemass_params}}
  \tablehead{
    \colhead{Filter} & \colhead{Rest-frame $\lambda_{\mathrm{rest}}$} & \colhead{Slope $\alpha$} & \colhead{Normalization $\log_{10}(A/\mathrm{kpc})$} & \colhead{Size at $5\times 10^{10}\ M_\odot$ ($A$)} & \colhead{Intrinsic Scatter $\sigma_{\log R_e}$} \\
    \colhead{} & \colhead{($\mu\mathrm{m}$)} & \colhead{} & \colhead{} & \colhead{(kpc)} & \colhead{(dex)}
  }
  \setlength{\tabcolsep}{12pt}
  \startdata
  F115W & $\sim 0.35$ & $0.51_{-0.12}^{+0.12}$ & $-0.06_{-0.04}^{+0.04}$ & $0.87_{-0.07}^{+0.08}$ & $0.18_{-0.03}^{+0.03}$ \\
  F150W & $\sim 0.45$ & $0.58_{-0.12}^{+0.13}$ & $-0.06_{-0.04}^{+0.04}$ & $0.86_{-0.08}^{+0.08}$ & $0.19_{-0.03}^{+0.03}$ \\
  F200W & $\sim 0.60$ & $0.59_{-0.12}^{+0.12}$ & $-0.08_{-0.04}^{+0.04}$ & $0.83_{-0.07}^{+0.08}$ & $0.17_{-0.02}^{+0.03}$ \\
  F277W & $\sim 0.85$ & $0.54_{-0.12}^{+0.12}$ & $-0.12_{-0.04}^{+0.04}$ & $0.77_{-0.06}^{+0.07}$ & $0.18_{-0.03}^{+0.03}$ \\
  F356W & $\sim 1.10$ & $0.56_{-0.13}^{+0.13}$ & $-0.15_{-0.04}^{+0.04}$ & $0.70_{-0.06}^{+0.07}$ & $0.19_{-0.03}^{+0.03}$ \\
  F444W & $\sim 1.35$ & $0.53_{-0.11}^{+0.13}$ & $-0.17_{-0.04}^{+0.04}$ & $0.68_{-0.06}^{+0.06}$ & $0.17_{-0.02}^{+0.03}$ \\
  \enddata
  \tablecomments{Best-fit parameters derived from Bayesian nested sampling with \texttt{dynesty} \citep{Speagle2020} for the sample of massive quiescent galaxies with $\log_{10}(M_*/M_\odot) > 10.3$ ($N = 24$) using COSMOS2020 quiescent SMF weights ($M_0 = 10^{10.49}\ M_\odot, \alpha_{\mathrm{SMF}} = -0.66$; \citealt{Weaver2023}). Values report the 50th percentile (median) with 16th and 84th percentile $1\sigma$ uncertainties.}
\end{deluxetable*}

\subsection{Spatially Resolved Analysis using \pixedfit} \label{sec:pixedfit_methods}

We perform spatially resolved stellar population and SED modeling across our sample using \texttt{piXedfit} \citep{Abdurrouf2021}. Firstly, we prepare multi-wavelength imaging cutouts across the NIRCam bands convolved to match the F444W PSF resolution to construct the datacube (Section~\ref{sec:psf_homogenization}). Then, we apply the SED-similarity-based pixel binning implemented in \texttt{piXedfit} \citep{Abdurrouf2017, Abdurrouf2021} to aggregate contiguous pixels with similar SED shapes, setting a target signal-to-noise ratio threshold of $\mathrm{SNR} \ge 5$ in the rest-frame optical/NIR broad bands while relaxing the SNR threshold to zero for filters with $\lambda_{\mathrm{rest}} < 4000\ \text{\AA}$ and for all medium bands \citep{Haryana_2025,Prasal2026}. After that, we fit the observed fluxes in each spatial bin using Bayesian inference with the Flexible Stellar Population Synthesis code \citep[\texttt{FSPS};][]{Conroy2009, Conroy2010} via \texttt{python-fsps} \citep{Foreman-Mackey2014}, adopting a \citet{Kroupa2001} initial mass function, Padova stellar evolutionary isochrones \citep{Girardi2000, Marigo2007, Marigo2008}, the MILES stellar library \citep{Sanchez-Blazquez2006, Falcon-Barroso2011}, a \citet{Calzetti2000} dust attenuation law, and a delayed-$\tau$ SFH. Following \citet{Haryana_2025}, we redistribute bin-level stellar mass and SFR to individual pixels weighted by pixel fluxes in F444W and F090W, respectively. From the resulting pixel-level parameter maps, we extract azimuthally averaged radial profiles of stellar mass surface density ($\Sigma_*$), SFR surface density ($\Sigma_{\mathrm{SFR}}$), sSFR, rest-frame colors, and assembly timescales in elliptical annuli. Specifically, we define the formation time ($t_{50}$) as the cosmic epoch when a given region reached 50\% of its total formed stellar mass, and the quenching epoch ($t_q$) as the cosmic time when the instantaneous SFR dropped to 10\% of its past time-averaged SFR, yielding the quenching timescale $\Delta t_q \equiv t_q - t_{50}$ \citep{Carnall2018, Nanayakkara2025}. We refer the reader to \citet{Prasal2026} for a comprehensive description of datacube construction, binning parameters, and SED modeling priors. Our choice of SFH does not change the results. The progression of quenching remains the same even when we use a double power law SFH.

\section{Results} \label{sec:results} \label{sec:sizemass}

We present our structural morphology measurements, multi-wavelength size--mass scaling relations, two-component decompositions, and spatially resolved stellar populations across our sample of 26 massive quiescent galaxies. Table~\ref{tab:sizemass_params} lists the best-fit size--mass parameters, and Table~\ref{tab:morph_summary_appendix} (Appendix~\ref{sec:appendix_morphology}) summarizes the multi-filter single-\sersic and bulge-disk decomposition parameters for the galaxies. Throughout our analysis, we report sample structural and morphological parameters as the median with 68\% confidence intervals derived from 1000 bootstrap iterations, whereas we report spatially resolved radial profiles and the \pysersic model residuals as the sample mean.

\subsection{Multi-Wavelength Size--Mass Scaling Relations and Cosmic Size Evolution} \label{sec:results_sizemass}

Our single-\sersic fits reveal moderately concentrated light profiles with median \sersic indices of $n \approx 2.9$ that remain stable from the rest-frame UV to near-infrared (see Table~\ref{tab:morph_summary_appendix}). The median projected axis ratio ($q \approx 0.6$) indicates flattened light distributions, consistent with oblate, disk-like structures. Across the six filters, the effective radius contracts monotonically toward longer wavelengths, decreasing from a median of $R_e = 0.99_{-0.18}^{+0.12}\ \mathrm{kpc}$ in F115W ($\lambda_{\mathrm{rest}} \sim 0.35\ \mu\mathrm{m}$) to $0.76_{-0.16}^{+0.14}\ \mathrm{kpc}$ in F444W ($\lambda_{\mathrm{rest}} \sim 1.35\ \mu\mathrm{m}$). Massive quiescent galaxies appear systematically smaller at longer wavelengths, consistent with negative color gradients where galaxy outskirts are bluer than their central cores \citep{Tortora2010,Wuyts2010,Guo2011,Szomoru2013, Mosleh2017, Suess2019, Suess2022, Clausen2025}.

\begin{figure*}[ht!]
  \centering
  \includegraphics[width=\textwidth]{ 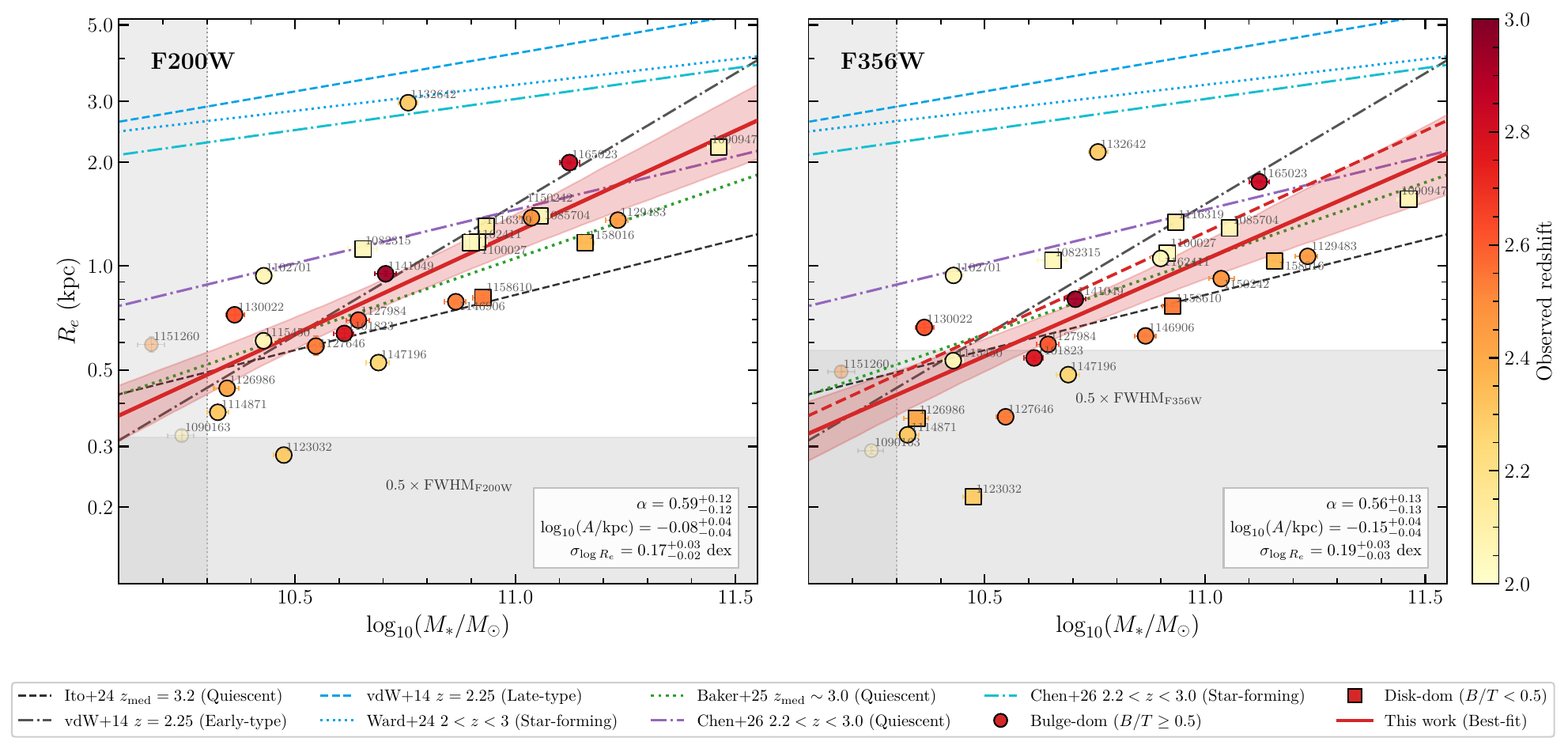}
  \caption{Size--mass relation for our sample of 26 massive quiescent galaxies at $z \sim 2.0$--$2.9$ in the MINERVA sample. Left: F200W filter ($\lambda_{\mathrm{rest}} \sim 0.60\ \mu\mathrm{m}$, rest-frame optical). Right: F356W filter ($\lambda_{\mathrm{rest}} \sim 1.10\ \mu\mathrm{m}$, rest-frame near-infrared). Circles represent bulge-dominated galaxies ($B/T \ge 0.5$), and squares represent disk-dominated galaxies ($B/T < 0.5$), color-coded by redshift, with individual galaxy IDs annotated. Two galaxies below our mass completeness threshold ($\log_{10}(M_*/M_\odot) < 10.3$, vertical grey shaded region) are shown with faint opacity. Solid red lines show our best-fitting power-law relations derived using \texttt{dynesty}, with shaded red regions indicating the $1\sigma$ intrinsic scatter ($\pm \sigma_{\log R_e}$). In the right panel, the red dashed line shows the F200W best-fitting relation for comparison. Literature quiescent relations are shown from \citet{vanderWel2014} ($z = 2.25$, black dash-dotted line), \citet{Ito2024} ($UVJ$-selected, median $z = 3.2$, black dashed line), \citet{Baker2025b} (median $z \approx 3.0$, F444W sizes, green dotted line), and \citet{Chen2026} at $2.2 < z < 3.0$ (purple dash-dotted line and purple points). Literature star-forming relations at comparable redshifts are shown from \citet{vanderWel2014} (blue dashed line), \citet{Ward2024} (light blue dotted line), and \citet{Chen2026} ($2.2 < z < 3.0$, cyan dashed line). The bottom horizontal shaded bands mark the empirical PSF resolution limits ($0.5 \times \mathrm{FWHM} = 0.32$ and $0.57\ \mathrm{kpc}$ in F200W and F356W, respectively, at median $z = 2.33$).}
  \label{fig:size_mass}
\end{figure*}

For galaxies above the mass threshold $\log_{10}(M_*/M_\odot) > 10.3$ ($N = 24$), we fit the size--mass relation across all six filters using \texttt{dynesty} (Figure~\ref{fig:size_mass}). In the rest-frame optical (F200W), we obtain a power-law slope of $\alpha = 0.59_{-0.12}^{+0.12}$, an intercept of $A = 0.83_{-0.07}^{+0.08}\ \mathrm{kpc}$ at pivot mass $M_* = 5 \times 10^{10}\ M_\odot$, and an intrinsic scatter of $\sigma_{\log R_e} = 0.17_{-0.02}^{+0.03}\ \mathrm{dex}$. In the rest-frame near-infrared (F356W), we derive a consistent slope of $\alpha = 0.56_{-0.13}^{+0.13}$, a smaller intercept of $A = 0.70_{-0.06}^{+0.07}\ \mathrm{kpc}$, and an intrinsic scatter of $\sigma_{\log R_e} = 0.19_{-0.03}^{+0.03}\ \mathrm{dex}$. The ratio between optical and near-infrared effective radii, $A_{\mathrm{F200W}} / A_{\mathrm{F356W}} \approx 1.2\times$, confirms that the mass profiles are more compact than the optical light profiles \citep{Suess2022, Martorano2024, Clausen2025, Kawinwanichakij2026}. Except for one galaxy (ID~1123032, with $R_e = 0.283\ \mathrm{kpc}$ and $n = 5.11$), all galaxies in our fitting sample comfortably exceed the empirical PSF half-resolution limit in F200W ($0.5\ \times \ \mathrm{FWHM} = 0.32\ \mathrm{kpc}$; Figure~\ref{fig:size_mass}). In F356W, the broader PSF ($0.5 \times \mathrm{FWHM} = 0.57\ \mathrm{kpc}$) places 7 of the 24 galaxies below this limit. Because \pysersic incorporates the wider PSF when fitting, their sizes remain constrained but less precise, so we regard F200W as our most robust size measurement. While \citet{Kawinwanichakij2026} identified an unresolved, diffused system (ZF-COS-20133, $n \sim 0.5$) where the fitting algorithm flattened the profile under PSF degeneracies, ID~1123032 exhibits a steep \sersic profile, reflecting a genuine, ultra-compact spheroidal core.

Our derived size--mass relations agree remarkably well with previous cosmic noon determinations for early-type and quiescent galaxies (Figure~\ref{fig:size_mass}). In the rest-frame optical (F200W) and near-infrared (F356W), our best-fit relations closely match the canonical early-type relation from CANDELS at $z = 2.25$ \citep[black dash-dotted line;][]{vanderWel2014} and the $UVJ$-selected quiescent relation at median $z = 3.2$ from \citet[black dashed line;][]{Ito2024}. In the rest-frame near-infrared (F356W), our relation also agrees with the size--mass relation of 18 spectroscopically confirmed quiescent galaxies from JADES at median $z \approx 3.0$ \citep[green dotted line;][]{Baker2025b}. \citet{Baker2025b} measured these sizes in F444W, which probes rest-frame $\sim 1.1\ \mu\mathrm{m}$ at $z \sim 3$, and caution that their relation is likely not statistically significant given the small sample. In comparison, the quiescent relation from \citet{Chen2026} at $2.2 < z < 3.0$ (purple dash-dotted line) exhibits slightly larger sizes at stellar masses ($\log_{10}(M_*/M_\odot) \lesssim 11.2$). This observed offset primarily stems from fundamental differences in sample selection. Whereas our study uses the sSFR\textsubscript{10} (10 Myr) as the quiescence criterion, \citet{Chen2026} isolated a quiescent sample using the sSFR\textsubscript{100} (100 Myr) criterion. Furthermore, we select our sample by leveraging 16 NIRCam medium band and broadband filters. However, \citet{Chen2026} selected quiescent candidates across the PRIMER field without medium-band coverage of MINERVA. Broad-band photometric selection without medium-band constraints could retain transition or green-valley systems with residual star-forming disks, which have sizes intermediate between those of star-forming and quiescent galaxies at fixed stellar mass \citep{EstradaCarpenter2023} and could elevate the average effective radius of their sample at intermediate masses. Nevertheless, both our sample and the literature quiescent relations remain sufficiently below the star-forming sequence as traced by \citealt{vanderWel2014, Ward2024, Chen2026}. Only one galaxy in our sample, ID~1132642 ($R_e \approx 3.0\ \mathrm{kpc}$ in F200W and $2.2\ \mathrm{kpc}$ in F356W), lies in this star-forming regime, possibly representing an extended disk-dominated outlier that has recently initiated quenching.

\begin{figure*}[ht!]
  \centering
  \includegraphics[width=\textwidth]{ 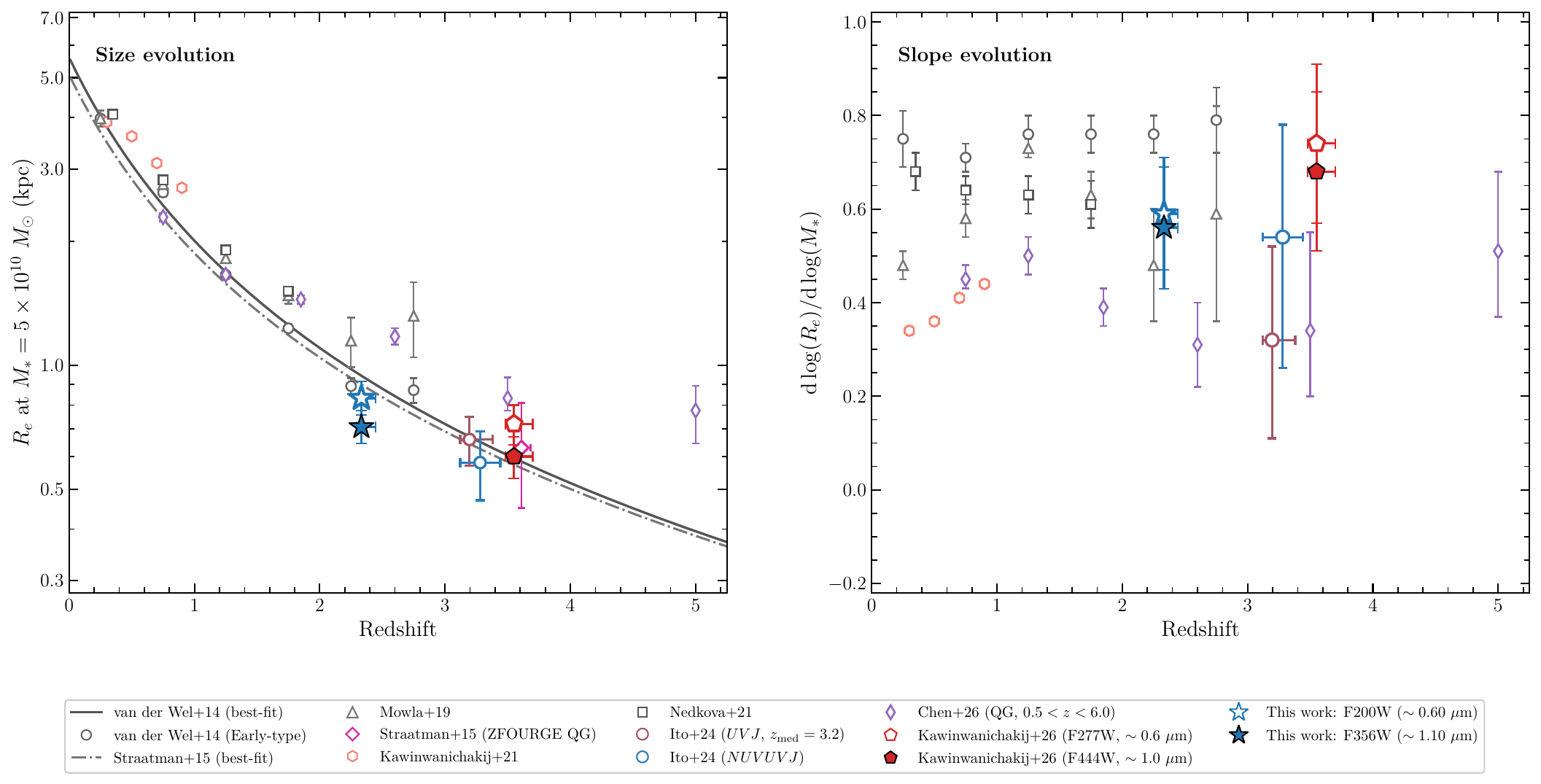}
  \caption{Evolution of the size--mass relation for massive quiescent galaxies from $z = 0$ to $z = 5$ compared with literature determinations. Left: Effective radius at pivot mass $M_* = 5 \times 10^{10}\ M_\odot$ as a function of redshift. Our MINERVA measurements at median redshift $z = 2.33_{-0.05}^{+0.12}$ are shown as the open blue star (F200W) and filled dark-blue star (F356W). Comparison quiescent samples include CANDELS early-type galaxies \citep[open black circles;][]{vanderWel2014}, COSMOS-DASH \citep[open grey triangles;][]{Mowla2019b}, HST WFC3 \citep[open grey squares;][]{Nedkova2021}, HSC-SSP \citep[open orange hexagons;][]{Kawinwanichakij2021}, ZFOURGE \citep[magenta diamonds;][]{Straatman2015}, JWST NIRCam at $z \sim 3.2$ \citep[UVJ brown circles and NUVUVJ blue circles;][]{Ito2024}, JWST NIRCam at $z \sim 3.5$ \citep[F277W open red pentagons and F444W filled dark red pentagons;][]{Kawinwanichakij2026}, and JWST PRIMER quiescent galaxies across $0.5 < z < 6.0$ \citep[purple open diamonds;][]{Chen2026}. Solid and dash-dotted curves show the rest-frame optical evolutionary tracks from \citet{vanderWel2014} ($R_e \propto (1+z)^{-1.48}$) and \citet{Straatman2015} ($R_e \propto (1+z)^{-1.44}$). Right: Evolution of the power-law slope $\alpha = \mathrm{d}\log(R_e)/\mathrm{d}\log(M_*)$ with redshift, demonstrating the persistent steep slope of quiescent systems up to $z \sim 4$. For our measurements, the vertical error bars show the 16th--84th percentiles of the \texttt{dynesty} posteriors, and the horizontal error bars show the 68\% confidence interval on the median redshift from 1000 bootstrap iterations. We derive the horizontal error bars of the \citet{Straatman2015}, \citet{Ito2024}, and \citet{Kawinwanichakij2026} points in the same way from their published redshifts, and all other error bars show the uncertainties reported in the original studies.}
  \label{fig:redshift_evolution}
\end{figure*}

At median redshift $z \approx 2.33$, our size--mass measurements follow canonical cosmic size evolution tracks [e.g., $r_e = 5.6(1+z)^{-1.48}\ \mathrm{kpc}$ from \citealt{vanderWel2014} and $r_e = 5.08(1+z)^{-1.44}\ \mathrm{kpc}$ from \citealt{Straatman2015}; Figure~\ref{fig:redshift_evolution}] bridging intermediate-redshift early-type galaxies \citep{Mowla2019b, Kawinwanichakij2021, Nedkova2021} and ultra-compact quiescent galaxies at $z > 3$ \citep{Ito2024, Kawinwanichakij2026}. Compared to local early-type galaxies of identical stellar mass, our characteristic sizes at $z = 2.33$ imply a $\sim 4$--$5\times$ growth in size from cosmic noon to the present day. In our rest-frame optical measurement in F200W ($R_e = 0.83_{-0.07}^{+0.08}\ \mathrm{kpc}$; open blue star), our characteristic size agrees within $1\sigma$ with both \citet{vanderWel2014} ($0.94\ \mathrm{kpc}$) and \citet{Straatman2015} ($0.90\ \mathrm{kpc}$). On the other hand, our rest-frame near-infrared measurement in F356W ($R_e = 0.70_{-0.06}^{+0.07}\ \mathrm{kpc}$, $\lambda_{\mathrm{rest}} \sim 1.10\ \mu\mathrm{m}$; filled dark-blue star) falls systematically $\approx 0.10\text{--}0.13\ \mathrm{dex}$ below these rest-frame optical tracks, reflecting negative color gradients and the intrinsic compactness of stellar mass profiles relative to rest-frame optical light \citep{Suess2019, Suess2022, Kawinwanichakij2026}. For comparison, in the rest-frame near-infrared ($1.5\ \mu\mathrm{m}$), \citet{Martorano2024} established an evolutionary relation of $R_e = 4.07(1+z)^{-1.29}\ \mathrm{kpc}$ for massive quiescent galaxies ($\log_{10}(M_*/M_\odot) \in [10.5, 11.0]$), which predicts $R_e \approx 0.86\ \mathrm{kpc}$ at $z = 2.33$. Our F356W measurement falls even further below this benchmark (by an additional $\approx 0.09\ \mathrm{dex}$), suggesting that our sample is more compact, although 7 of the 24 galaxies lie below the F356W resolution limit. Furthermore, as shown in the right panel of Figure~\ref{fig:redshift_evolution}, the quiescent size--mass power-law slope shows no systematic evolution from $z \sim 4$ down to $z \sim 0$. Most determinations lie at $\alpha \approx 0.45$--$0.8$ \citep{vanderWel2014, Mowla2019b, Nedkova2021, Kawinwanichakij2026}, whereas samples extending to lower masses or using different selections, such as the $UVJ$-selected sample of \citet{Ito2024}, yield shallower slopes of $\alpha \approx 0.3$--$0.45$ \citep{Kawinwanichakij2021, Chen2026}. Our MINERVA measurements demonstrate that massive quiescent galaxies already obeyed a steep size dependence on stellar mass ($R_e \propto M_*^{\sim 0.6}$) by cosmic noon, matching local early-type galaxies.

\subsection{Two-Component Bulge--Disk Decomposition} \label{sec:results_bulgedisk}

In F200W, two-component decompositions with \texttt{sersic\_exp} reveal that roughly two-thirds ($N=18$) of our galaxies are bulge-dominated ($B/T \ge 0.5$), while approximately one-third ($N = 8$) retain dominant exponential disks ($B/T < 0.5$) with median $B/T = 0.59_{-0.06}^{+0.04}$ in F200W and $0.57_{-0.05}^{+0.05}$ in F356W; Table~\ref{tab:morph_summary_appendix}.

\begin{figure*}[ht!]
  \centering
  \includegraphics[width=\textwidth]{ 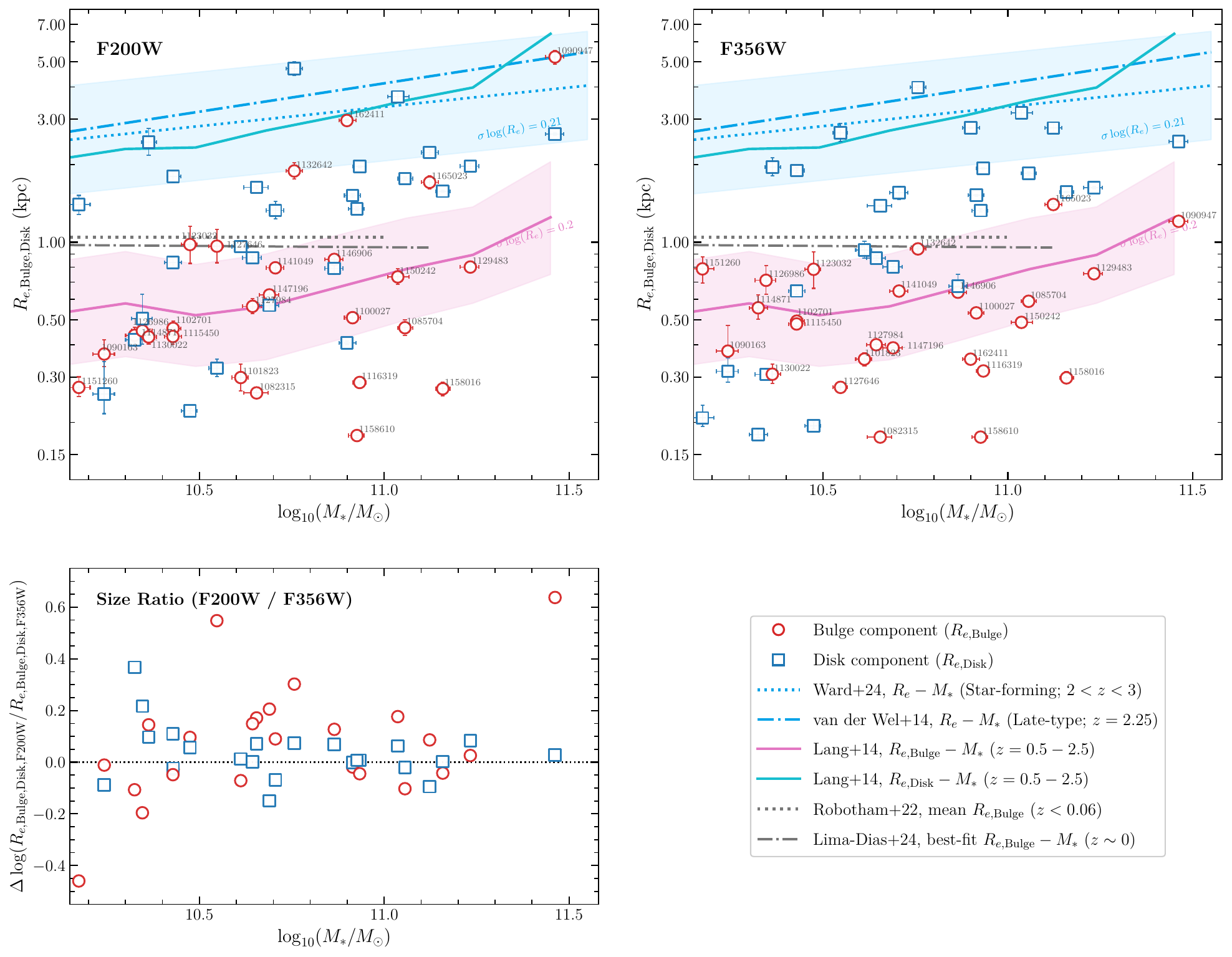}
  \caption{Two-component bulge and disk scaling relations for our sample of massive quiescent galaxies. Top-left: Effective radius of bulge components ($R_{e,\mathrm{Bulge}}$, red open circles) and disk components ($R_{e,\mathrm{Disk}}$, blue open squares) in F200W vs.\ total stellar mass, with galaxy IDs annotated. Top-right: Bulge and disk effective radii in F356W vs.\ total stellar mass. Comparison curves show the $z \sim 2.25$ late-type relation from \citet{vanderWel2014} (blue dash-dotted line), the $2 < z < 3$ star-forming disk relation from \citet{Ward2024} (blue dotted line and shaded envelope), the $0.5 < z < 2.5$ quiescent bulge and disk relations from \citet{Lang2014} (pink and cyan curves, with the pink shaded envelope showing the $1\sigma$ scatter of the bulge relation), and local bulge relations from \citet{Robotham2022} (grey dotted line) and \citet{LimaDias2024} (grey dash-dotted line). Bottom-left: Logarithmic size ratio between F200W and F356W for bulges (red circles) and disks (blue squares) vs.\ stellar mass, relative to the dotted zero line.}
  \label{fig:bulge_disk}
\end{figure*}

The decoupled bulges exhibit steep light profiles with median \sersic indices of $n_{\mathrm{bulge}} = 3.77_{-0.09}^{+0.44}$ in F200W and $3.37_{-0.37}^{+0.21}$ in F356W. Across the entire mass range ($\log_{10}(M_*/M_\odot) \approx 10.3$--$11.6$), bulges remain ultra-compact, with a median radius of $R_{e,\mathrm{Bulge}} \approx 0.49\ \mathrm{kpc}$ in both bands (Figure~\ref{fig:bulge_disk}). As shown in Figure~\ref{fig:bulge_disk}, bulge sizes exhibit noticeably tighter dispersion in F356W ($\sigma_{\log R_e} = 0.22\ \mathrm{dex}$) than in F200W ($\sigma_{\log R_e} = 0.34\ \mathrm{dex}$), where residual star-forming clumps and dust attenuation introduce larger decomposition uncertainties. Moreover, bulge sizes remain remarkably constant across the entire sample (median $R_{e,\mathrm{Bulge}} = 0.53_{-0.04}^{+0.15}\ \mathrm{kpc}$ at $\log M_* \le 10.5$ and $0.44_{-0.07}^{+0.12}\ \mathrm{kpc}$ at $\log M_* > 10.5$), exhibiting negligible dependence on total stellar mass. This mass invariance of central bulges agrees with findings across both local galaxies \citep{Robotham2022, LimaDias2024} and intermediate-redshift samples \citep{Lang2014, Nedkova2024}. Our bulge measurements closely trace the $0.5 < z < 2.5$ quiescent bulge relation from resolved mass maps in \citet{Lang2014} ($R_{e,\mathrm{Bulge}} \approx 0.53\text{--}0.70\ \mathrm{kpc}$). Compared to local bulges ($R_{e,\mathrm{Bulge}} \approx 1.0\ \mathrm{kpc}$; \citealt{Robotham2022, LimaDias2024}), our cosmic-noon bulges are only $\sim 2\times$ smaller, whereas the total sizes of quiescent galaxies at fixed stellar mass grow by $\sim 4$--$5\times$ over the same cosmic epoch \citep{vanderWel2014}. This contrast suggests that central compact cores undergo modest structural change following early compaction, consistent with inside-out growth, in which subsequent mass accretion predominantly builds the outer envelope \citep{vanDokkum2010, Lang2014, Nedkova2024}.

Conversely, the surrounding disks extend $\approx 3\times$ further, with median sizes of $R_{e,\mathrm{Disk}} = 1.38_{-0.28}^{+0.20}\ \mathrm{kpc}$ in F200W and $1.54_{-0.21}^{+0.17}\ \mathrm{kpc}$ in F356W. For the disk-dominated subset ($B/T < 0.5$), disk sizes have a median of $R_{e,\mathrm{Disk}} = 1.60_{-0.09}^{+0.16}\ \mathrm{kpc}$ in F200W and $1.52_{-0.16}^{+0.05}\ \mathrm{kpc}$ in F356W. These disks lie $\sim 0.3$--$0.4$ dex below the star-forming galaxy size--mass relations at cosmic noon \citep{vanderWel2014, Ward2024}, beyond their $\sim 0.2$ dex intrinsic scatter. This agrees with \citet{Nedkova2024}, who found that quiescent disks are typically smaller than star-forming disks at fixed stellar mass. Crucially, both bulges and disks show negligible size changes between F200W and F356W ($\Delta \log R_e \le 0.04\ \mathrm{dex}$; bottom-left panel of Figure~\ref{fig:bulge_disk}). This implies that the global color gradient and wavelength-dependent size contraction observed in single-\sersic fits stem from the increased flux dominance of the compact bulge at longer wavelengths, rather than the radial shrinkage of individual components.

\begin{figure*}[ht!]
  \centering
  \includegraphics[width=\textwidth]{ 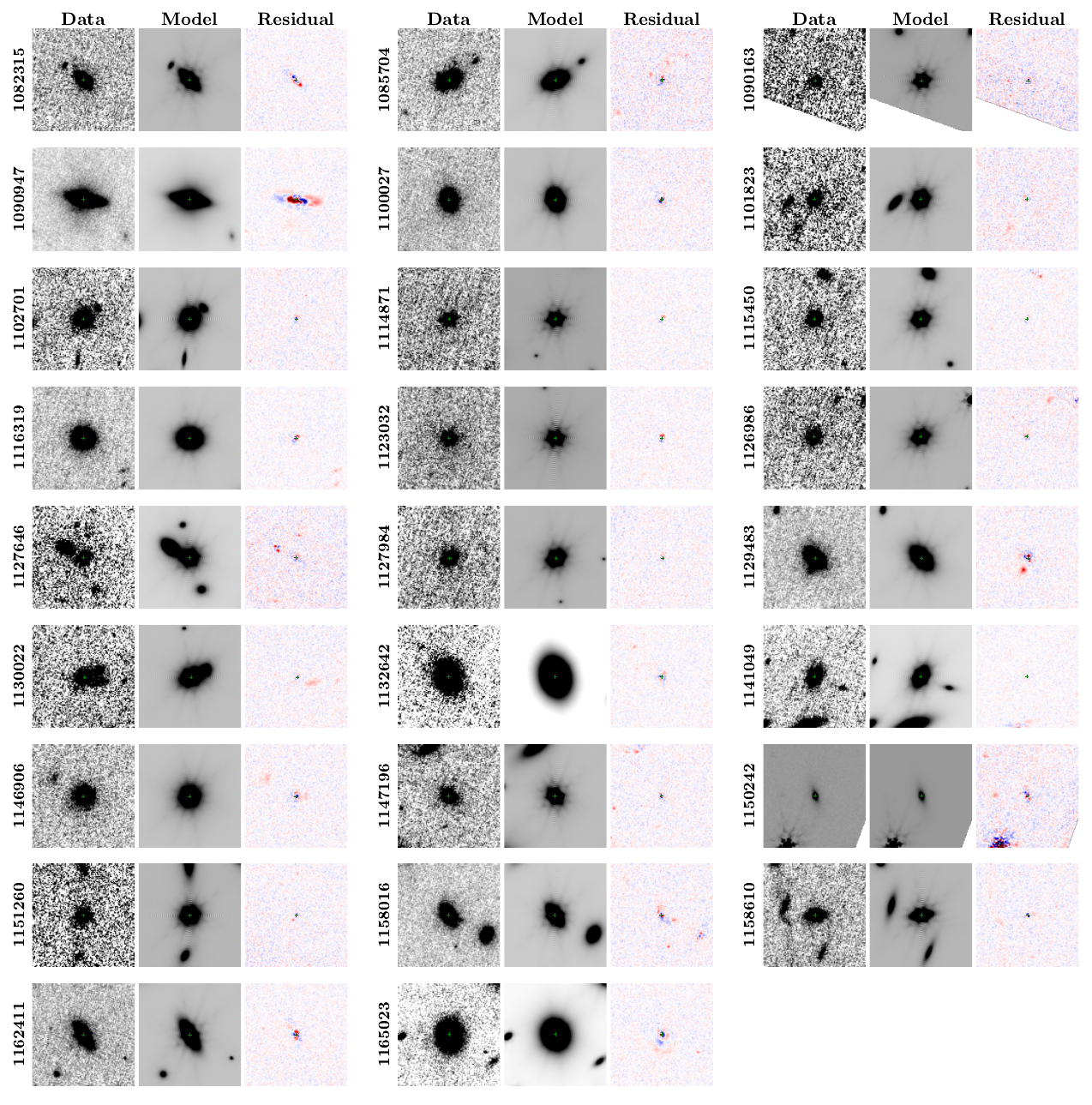}
  \caption{Two-component \pysersic fits for the galaxies in our sample in the F200W filter (4\arcsec\ cutouts). The panels display the original images, the best-fit models, and the resulting residual maps. The color scale limits (v\textsubscript{min} and v\textsubscript{max}) for the residual plots are set to $\pm1$ (in 10 nJy), where blue regions indicate less flux than the model and red regions indicate excess flux. The average absolute residual is 4.66\% $\pm$ 1.14\%, with values ranging from 0.01\% to 29.12\%. The standard deviation of the mean is calculated via bootstrapping for 1000 iterations.}
  \label{fig:pysersic_residual_f200w}
\end{figure*}

\begin{figure*}[ht!]
  \centering
  \includegraphics[width=\textwidth]{ 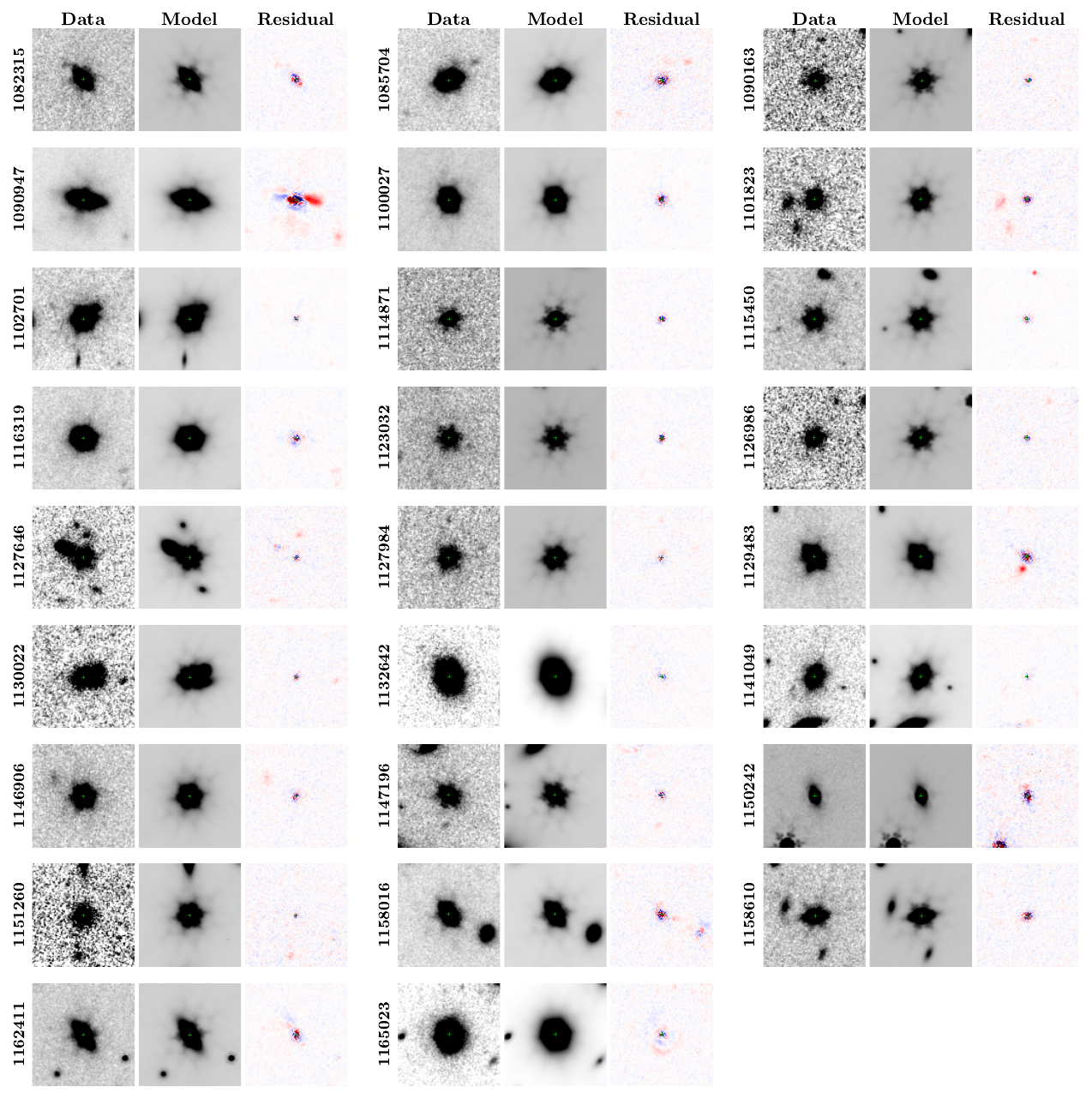}
  \caption{Two-component \pysersic fits for the galaxies in our sample in the F356W filter (4\arcsec\ cutouts). The panels display the original images, the best-fit models, and the resulting residual maps. The color scale limits (v\textsubscript{min} and v\textsubscript{max}) for the residual plots are set to $\pm1$ (in 10 nJy), where blue regions indicate less flux than the model and red regions indicate excess flux. The average absolute residual is 1.78\% $\pm$ 0.31\%, with values ranging from 0.09\% to 7.34\%. The standard deviation of the mean is calculated via bootstrapping for 1000 iterations.}
  \label{fig:pysersic_residual_f356w}
\end{figure*}

Figures~\ref{fig:pysersic_residual_f200w} and \ref{fig:pysersic_residual_f356w} display the cutouts, best-fit models, and residual maps for our sample in F200W and F356W, respectively. The two-component models reproduce the observed light distributions well, yielding low average absolute residuals of $4.66\% \pm 1.14\%$ in F200W and $1.78\% \pm 0.31\%$ in F356W.

\subsection{Spatially Resolved Analysis and Radial Profiles} \label{sec:results_resolved}

\begin{figure*}[ht!]
  \centering
  \includegraphics[width=0.9\textwidth]{ 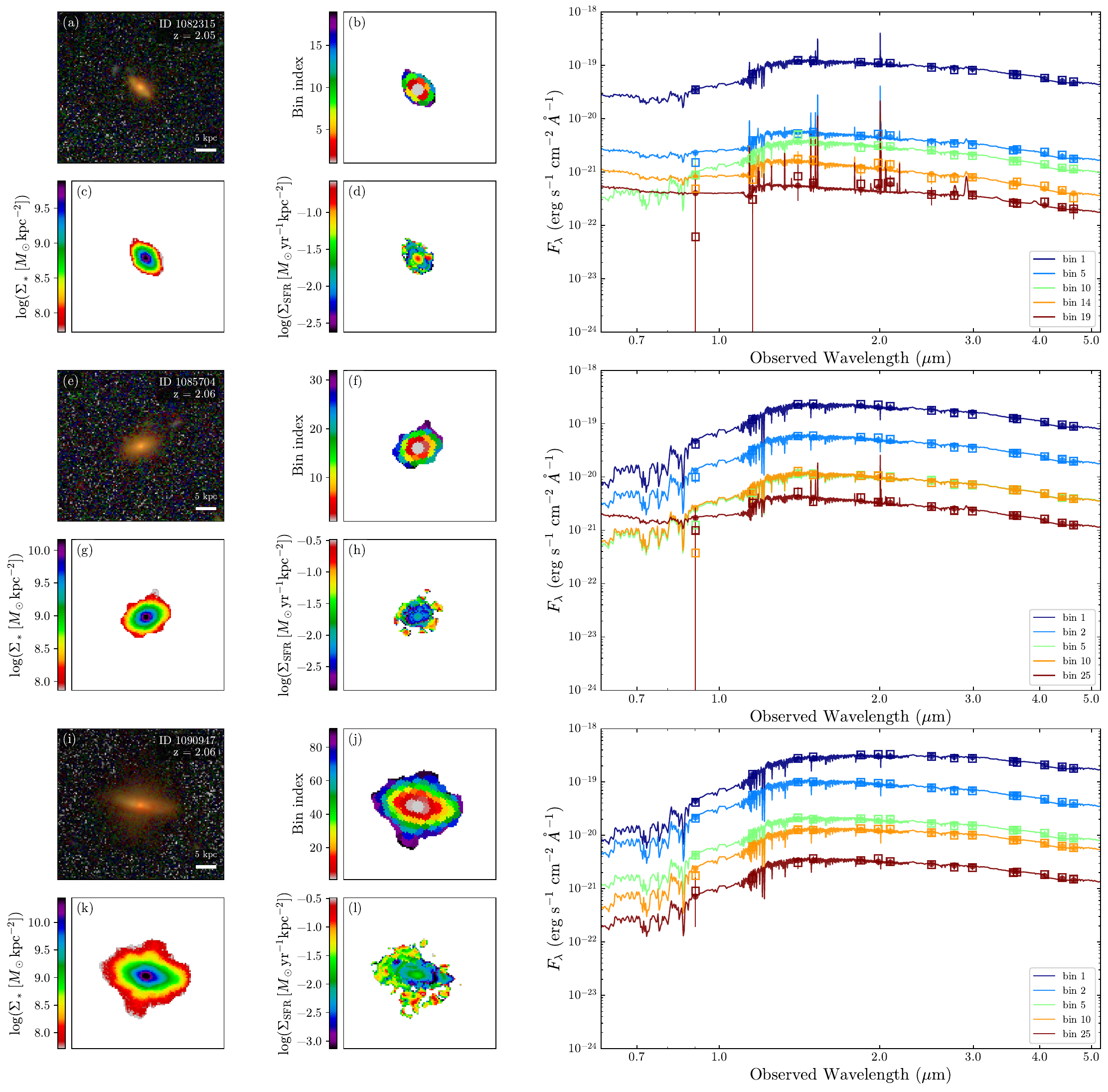}
  \caption{Spatially resolved SED fits for three representative galaxies in our sample (\texttt{ID} = 1082315 at $z=2.05$, 1085704 at $z=2.06$, and 1090947 at $z=2.06$). Panels (a), (e), and (i) are the RGB images (using F200W, F150W, and F115W filters), (b), (f), and (j) are the bin index maps, (c), (g), and (k) are the pixel-level stellar mass maps, and (d), (h), and (l) are the pixel-level SFR maps. The panels on the right show the spectra of different bins for these three galaxies. Filled circles are the model fluxes, and the open squares are the observed fluxes, with error bars. The solid lines are the best-fit SEDs for each bin.}
  \label{fig:haryana_style_figure}
\end{figure*}

Using \texttt{piXedfit} \citep{Abdurrouf2021}, we reconstruct spatially resolved maps of stellar mass surface density ($\Sigma_*$) and SFR surface density ($\Sigma_{\mathrm{SFR}}$). Figure~\ref{fig:haryana_style_figure} illustrates representative \pixedfit SED fits for three galaxies in our sample (\texttt{ID} = 1082315, 1085704, and 1090947).

\begin{figure*}[ht!]
  \centering
  \includegraphics[width=0.85\textwidth]{ 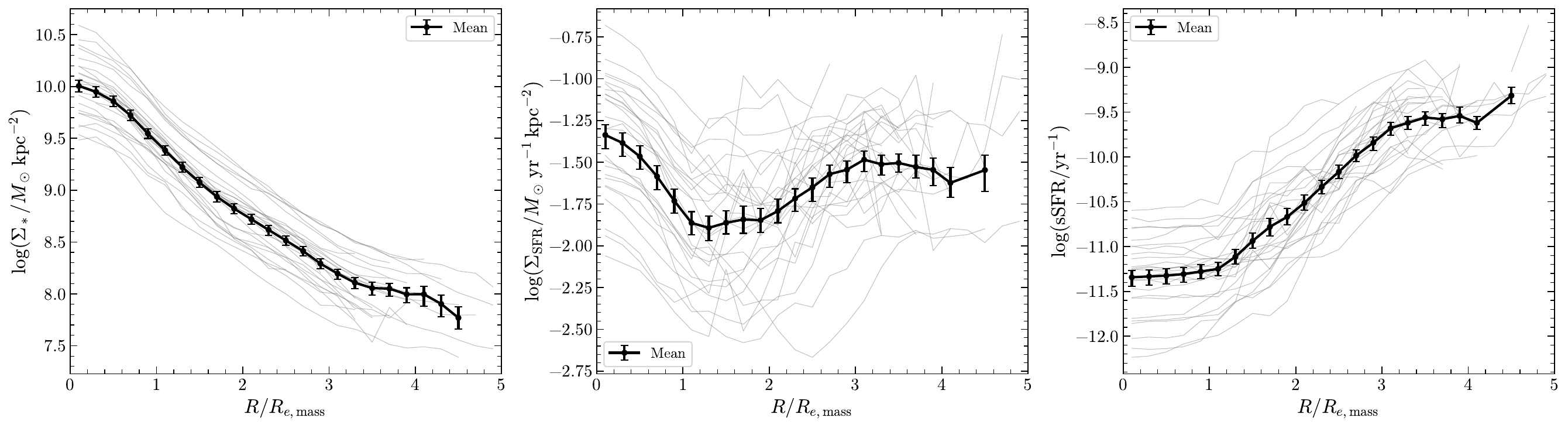}
  \includegraphics[width=0.85\textwidth]{ 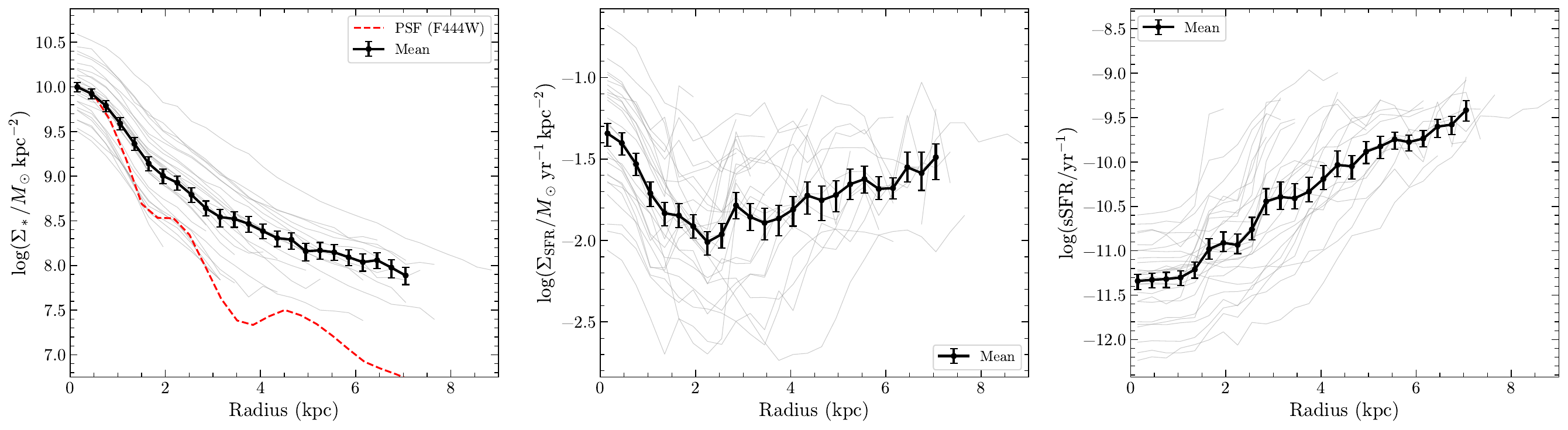}
  \caption{Mean radial profiles for the stellar mass, SFR, and sSFR for our sample of 26 galaxies. The top panels show profiles normalized by the half mass radius ($R_e$), following \citet{Laishram2026}. The bottom panels show the profiles with the radius $R$ in increments of 0.3 kpc for direct comparison with \citet{Haryana_2025}. Left: Stellar mass radial profiles. Middle: SFR radial profiles. Right: sSFR radial profiles. The red dashed line in the stellar mass radial profile plot in the bottom panel is the PSF profile for the F444W filter, the reference filter for PSF matching. We can observe an increasing sSFR gradient, indicating inside-out quenching. The error bars are the 68\% confidence intervals of the mean, derived from bootstrapping for 1000 iterations.}
  \label{fig:radial_profiles}
\end{figure*}

Figure~\ref{fig:radial_profiles} presents the mean radial profiles of $\Sigma_*$, $\Sigma_{\mathrm{SFR}}$, and $\mathrm{sSFR}$ across our sample. We normalize the radial profiles by the half-mass radius ($R/R_{e,\mathrm{mass}}$) following \citet{Laishram2026}, and plot them against physical radius in kpc following \citet{Haryana_2025}. We find the mean half-mass radius of our sample as $R_{e,\mathrm{mass}} = 1.29 \pm 0.06\ \mathrm{kpc}$. Because we derive $R_{e,\mathrm{mass}}$ from \pixcomma\ maps at the F444W resolution ($\mathrm{FWHM} \approx 1.3\ \mathrm{kpc}$) without PSF deconvolution, $R_{e,\mathrm{mass}}$ is PSF-broadened and exceeds the PSF-corrected \sersic sizes (Section~\ref{sec:results_sizemass}), so we use it only to normalize the radial profiles.

The stellar mass surface density decreases by $\approx 2.1\ \mathrm{dex}$ from $R \lesssim 0.3\ \mathrm{kpc}$ to $R \approx 7\ \mathrm{kpc}$, extending well beyond the F444W PSF profile. The mean profile implies a core stellar mass surface density of $\log \Sigma_{1\mathrm{kpc}} \approx 10\ M_\odot\,\mathrm{kpc}^{-2}$. The SFR surface density decreases till $R \approx 2 \ \rm{kpc}$, and then increases towards the outskirts. In contrast, the specific star formation rate is strongly suppressed in the central core and rises continuously outward by two orders of magnitude till $R \approx 7\ \mathrm{kpc}$. This steep positive sSFR gradient provides direct empirical evidence for inside-out quenching operating across the sample. Similar suppression of central sSFR by 1--2 orders of magnitude with actively star-forming outskirts has been observed in previous studies at cosmic noon \citep{Haryana_2025, Laishram2026,Prasal2026}, reflecting elevated central mass-to-light ratios and passive cores, while star formation continues in outer disks.

\begin{figure}[ht!]
  \centering
  \includegraphics[width=0.85\columnwidth]{ 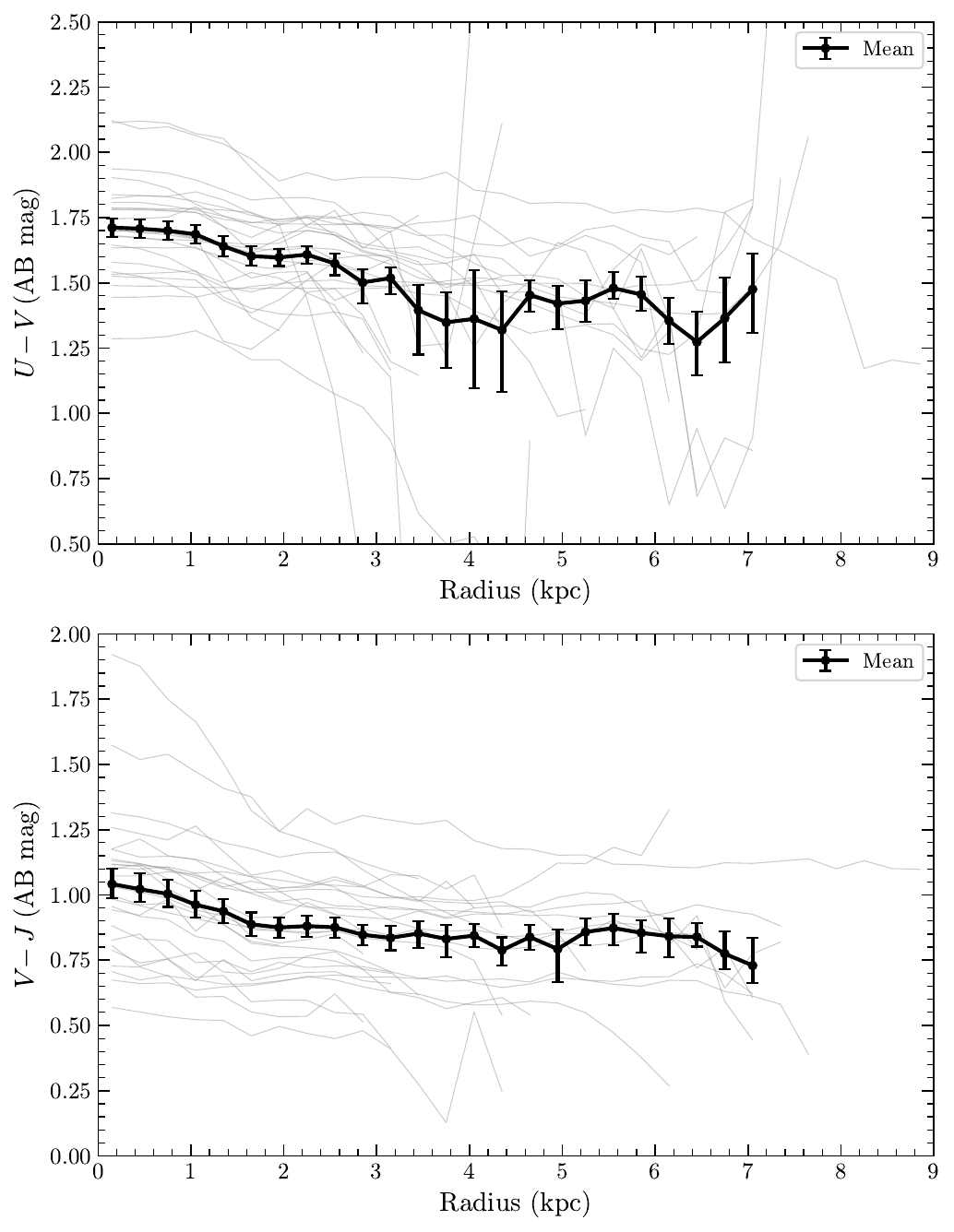}
  \caption{Mean radial profiles of the $U-V$ and $V-J$ colors as a function of radius in kpc for our sample of 26 galaxies. The error bars are the 68\% confidence intervals of the mean, derived from bootstrapping for 1000 iterations.}
  \label{fig:uvj_radial_profiles}
\end{figure}

Furthermore, radial $U-V$ and $V-J$ color profiles (Figure~\ref{fig:uvj_radial_profiles}; \citealt{Maiz-Apellaniz2006, Skrutskie2006}; see \citealt{Haryana_2025,Prasal2026} for the exact methodology) display negative radial gradients, declining from red central cores to bluer outskirts. These gradients confirm that passively evolving, older stellar populations dominate the galactic cores, while younger stellar populations populate the outskirts \citep[][see also \citealt{Haryana_2025, Laishram2026, Cutler2026}]{Suess2019, Suess2022}.

\begin{figure}[ht!]
  \centering
  \includegraphics[width=1\columnwidth]{ 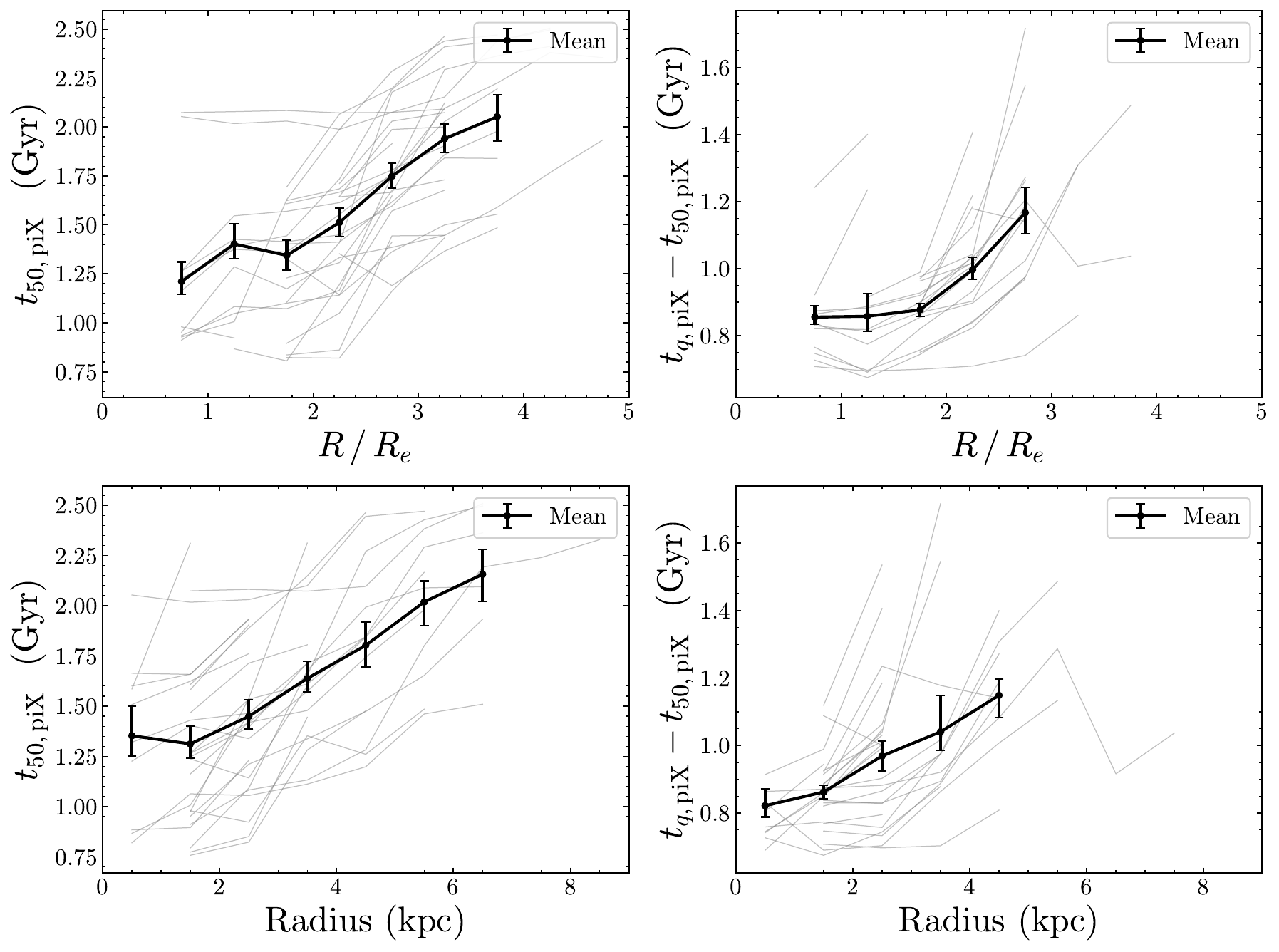}
  \caption{Mean radial profiles of the formation time ($t_{50, \,\mathrm{piX}}$, on the left; cosmic time when the region reached 50\% of its total formed stellar mass) and the quenching timescale ($t_{q, \,\mathrm{piX}}- t_{50, \,\mathrm{piX}}$, on the right; elapsed duration between 50\% mass assembly and when the SFR drops to 10\% of its past time-averaged value) for our sample of 26 galaxies. The top panels show the profile normalized by the half mass radius ($R_e$) in bins of 0.5 $R_e$, while the bottom panels show the profile with radius in kpc in bins of 1 kpc. The error bars are the 68\% confidence intervals of the mean, derived from bootstrapping for 1000 iterations.}
  \label{fig:tau_radial_profiles}
\end{figure}

Radial profiles of the formation time ($t_{50, \,\mathrm{piX}}$) and quenching timescale ($t_{q, \,\mathrm{piX}} - t_{50, \,\mathrm{piX}}$) provide further evidence for this inside-out progression (Figure~\ref{fig:tau_radial_profiles}). The $t_{50, \,\mathrm{piX}}$ profile displays a positive gradient, with the inner regions reaching 50\% of their stellar mass at least $\approx 0.7\ \rm{Gyr}$ earlier than the outer regions. Similarly, the quenching timescale is shortest in the inner regions and lengthens toward the outskirts, demonstrating that star formation ceased earliest and most rapidly in the galaxy centers \citep{Haryana_2025}.

\section{Discussion} \label{sec:discussion}

\subsection{Cosmic Size Evolution and Two-Phase Galaxy Assembly} \label{sec:disc_evolution}

At pivot mass $M_* = 5 \times 10^{10}\ M_\odot$, our derived characteristic effective radii ($A = 0.83_{-0.07}^{+0.08}\ \mathrm{kpc}$ in F200W and $0.70_{-0.06}^{+0.07}\ \mathrm{kpc}$ in F356W) and steep power-law slopes ($\alpha = 0.59 \pm 0.12$ and $0.56 \pm 0.13$) show remarkable agreement with previous size--mass scaling relations established for early-type and quiescent galaxies at cosmic noon \citep[e.g.,][]{vanderWel2014, Ito2024, Baker2025b}. Massive quiescent galaxies at $z \sim 2.3$ are a factor of $4\text{--}5$ smaller than early-type galaxies of comparable stellar mass in the local universe. Despite this global compactness, our two-component decomposition reveals that while their central bulges ($R_{e,\mathrm{Bulge}} \approx 0.49\ \mathrm{kpc}$) are more compact than present-day galaxy bulges \citep{Robotham2022, LimaDias2024}, their central core densities ($\log\Sigma_{1\mathrm{kpc}} \approx 10.0\ M_\odot\,\mathrm{kpc}^{-2}$) already match the dense inner cores of local quiescent galaxies \citep{Fang2013}.

This structural continuity supports the canonical two-phase formation model of massive galaxies \citep{Naab2009, Oser2010, vanDokkum2010, Oser2012}. In the first phase ($z > 2$), rapid dissipative gas compaction and inside-out quenching assemble the compact stellar core \citep{Dekel2014, Zolotov2015, Tacchella2016}. Recent spatially resolved JWST analyses support this picture and found that the inner $1\,\mathrm{kpc}$ cores of massive quiescent galaxies formed early during compact, gas-rich starbursts, reaching high central stellar mass densities and quenching rapidly from the inside out while their outskirts remain extended \citep{Haryana_2025, Laishram2026}. In the second phase ($z < 2$), star formation remains largely dormant, and subsequent size and mass growth proceed primarily through dissipationless (dry) minor mergers \citep{Naab2009, vanDokkum2010, Oser2012, Haryana_2025}. Because minor mergers deposit accreted stars predominantly in the diffuse outer halo, individual galaxies expand along steep structural tracks ($ R_e \propto M_*^ 2$) while leaving the dense central core dynamically undisturbed \citep{Bezanson2009, vanDokkum2010, Hilz2013, Haryana_2025}. This scenario is consistent with a size--mass slope that does not evolve with time. If minor mergers grow galaxies of all masses by a similar factor, the relation shifts to larger sizes while keeping its slope. As illustrated in the right panel of Figure~\ref{fig:redshift_evolution}, the high-mass quiescent size--mass slope shows no systematic evolution from $z \approx 4$ to $z = 0$, with most determinations at $\alpha \approx 0.45$--$0.8$ and shallower values ($\alpha \approx 0.3$--$0.45$) from samples extending to lower masses or using different selections \citep[e.g.,][]{Mowla2019b, Kawinwanichakij2021, Nedkova2021, Ito2024, Chen2026, Kawinwanichakij2026}, consistent with our MINERVA measurements.

Our steep best-fit slope ($\alpha = 0.59 \pm 0.12$ in F200W, $0.56 \pm 0.13$ in F356W) agrees with canonical early-type scaling at cosmic noon and higher redshifts \citep[e.g., $\alpha \sim 0.7\text{--}0.8$;][]{vanderWel2014, Kawinwanichakij2026}, but exceeds shallower slopes reported in some photometric surveys (e.g., $\alpha = 0.32 \pm 0.20$ for the $UVJ$-selected sample of \citealt{Ito2024}). The stricter $NUVUVJ$-selected sample of \citet{Ito2024} yields $\alpha = 0.54_{-0.28}^{+0.24}$, which aligns better with our measurements. As demonstrated by \citet{Kawinwanichakij2026}, this disparity primarily stems from sample selection and stellar mass thresholding. When adopting a strict high-mass cut ($\log_{10}(M_*/M_\odot) > 10.3$), the quiescent size--mass relation remains steep ($\alpha \approx 0.74 \pm 0.20$). However, when lower-mass galaxies or systems with substantial stellar mass uncertainties scatter into the sample, the fit flattens to $\alpha \sim 0.36 \pm 0.20$ \citep{Kawinwanichakij2026}, implying that such low-mass inclusion naturally biases single power-law fits toward shallower slopes. Our slope does not show this flattening on including those galaxies because our two galaxies below the mass cut lie within the intrinsic scatter on opposite sides of the best-fitting relation (Section~\ref{sec:sizemass_fitting}).

While \citet{Chen2026} found larger average sizes for quiescent galaxies at intermediate masses ($\log_{10}(M_*/M_\odot) \lesssim 11.2$), this offset predominantly reflects differences in sample selection and the effect of medium bands. When restricted to robustly selected quenched samples \citep[][this work]{vanderWel2014, Ito2024, Baker2025b}, quiescent systems at $z \sim 2$--$3$ consistently follow a steep scaling relation that lies $\approx 0.6\ \mathrm{dex}$ below the star-forming sequence at the pivot mass, with the offset decreasing toward the highest masses (Figure~\ref{fig:size_mass}). This structural offset relative to star-forming galaxies suggests that massive galaxies become more compact before or during quenching, although it could also partly reflect progenitor bias, since quiescent galaxies at $z \sim 2$--$3$ likely descend from star-forming progenitors that were already smaller at earlier epochs.

The best-fit intrinsic scatter of our sample ($\sigma_{\log R_e} = 0.17_{-0.02}^{+0.03}\ \mathrm{dex}$ in F200W and $0.19_{-0.03}^{+0.03}\ \mathrm{dex}$ in F356W) aligns with literature determinations at $z \sim 2.5$--$3.5$ \citep[$\sigma \approx 0.14$--$0.28\ \mathrm{dex}$;][]{vanderWel2014, Ito2024, Kawinwanichakij2026}. This structural dispersion demonstrates that size is not uniquely determined by stellar mass alone. Instead, it reflects diverse evolutionary pathways shaped by progenitor populations \citep[e.g.,][]{Carnall2023b, Kawinwanichakij2026}, distinct compaction rates \citep[fast vs.\ slow; e.g.,][]{Zolotov2015}, and local environments \citep[e.g.,][]{Matharu2019, Ito2023, Kawinwanichakij2026}.

Instrumental capabilities also introduce systematic differences between JWST and earlier HST studies. Earlier canonical scaling relations at cosmic noon \citep[e.g.,][]{vanderWel2014, Straatman2015} relied on HST/WFC3 F160W imaging with a comparatively broad PSF ($\mathrm{FWHM} \sim 0\farcs18$--$0\farcs20$). For compact galaxies with $R_e \lesssim 0.5 \times \mathrm{FWHM}_{\mathrm{HST}}$, PSF smearing strongly limits the precision of HST size measurements. In contrast, JWST/NIRCam provides a significantly sharper PSF ($\mathrm{FWHM} \sim 0\farcs07$--$0\farcs14$) and heightened sensitivity. These improvements enable NIRCam to resolve compact galactic cores and their internal bulge components without artificial smoothing, while simultaneously detecting faint extended light profiles at large radii.

Additionally, the systematic size contraction across the NIRCam bands ($\approx 22\%$ from F115W to F444W, yielding an offset of $\approx 0.07\ \mathrm{dex}$ in normalization $A$ between F200W and F356W) aligns with negative rest-frame color gradients across our galaxies. In this picture, rest-frame optical light captures extended, younger stellar populations in the outskirts, while rest-frame near-infrared light traces the dense, older stellar mass core, consistent with literature findings of radial mass-to-light ratio gradients \citep[][see also \citealt{Kawinwanichakij2026}]{Suess2019, Suess2022, Martorano2024}.

\subsection{Inside-Out Quenching, Dissipative Compaction, and Disk Survival} \label{sec:disc_compaction}

The steep positive sSFR gradient, accompanied by negative rest-frame color gradients and older central stellar populations, provides robust empirical confirmation of inside-out quenching across our sample \citep[][see also \citealt{Haryana_2025, Laishram2026}]{Tacchella2015, Tacchella2016, Lin2019, Nelson2021}. Spatially resolved age radial profiles describe this inside-out trajectory, where the inner regions ($R < 4\ \mathrm{kpc}$) formed significantly earlier and quenched much faster than the extended outskirts. These findings support the dissipative gas compaction framework \citep{Dekel2014, Zolotov2015, Tacchella2016, Barro2017}. In this picture, violent disk instabilities or gas-rich mergers at $z > 2$ funnel cold gas toward the galaxy center, igniting a compact starburst (the ``blue nugget'' phase) that rapidly builds the dense stellar core. Once the core attains maximum central compactness, nuclear star formation terminates abruptly, setting up the structural conditions for subsequent feedback and dynamical stabilization.

These compact bulges confirm that central structural concentration couples tightly to star formation cessation, both at cosmic noon \citep{Whitaker2015, Barro2017, Laishram2026,Prasal2026} and in the local Universe \citep{Bait2017, Pan2018}. Importantly, our galaxies do not conform to purely pressure-supported spherical geometries. The sample exhibits flattened projected axis ratios ($q \approx 0.6$), and approximately one-third of them retain prominent exponential disks ($B/T < 0.5$). The survival of extended disk components demonstrates that quenching does not universally destroy disk morphology, consistent with kinematic and structural evidence that many high-redshift quiescent galaxies are rotationally supported disks or fast rotators rather than dispersion-dominated spheroids \citep{vanderWel2014, Toft2017, Newman2018, Scholtz2026}.

\subsection{Physical Quenching Drivers} \label{sec:disc_quenching}

Dissipative compaction, AGN feedback, and morphological stabilization may act together as complementary stages in the quenching of massive galaxies at cosmic noon. In this scenario, compaction first assembles the dense stellar core and channels gas toward the central supermassive black hole \citep{Hopkins2008, Barro2017}. Once the black hole grows sufficiently massive, AGN feedback in the form of radiation pressure or kinetic winds can heat and expel the remaining nuclear gas \citep{Croton2006, Fabian2012, Nelson2019, Nelson2021}. Such central gas removal would rapidly shut down star formation in the core, consistent with the short central quenching timescale and the positive sSFR gradient that we observe.

While central feedback halts star formation in the nucleus, it need not destroy the outer disk. The survival of extended, flattened disk components in our sample agrees with JWST kinematic evidence that massive quiescent galaxies at cosmic noon commonly remain rotationally supported \citep{Slob2025}. It is also consistent with morphological quenching, in which a massive bulge stabilizes the remaining gas disk against fragmentation \citep{Martig2009, Genzel2014}, although testing this requires gas and kinematic measurements that are beyond the scope of this work.

Together, these processes provide a plausible explanation for our observations, although imaging alone cannot isolate their individual contributions. To quantify the strength of any potential AGN contribution, we compute the \pysersic residuals for the central region of $0\farcs2\times0\farcs2$ (Figures~\ref{fig:pysersic_residual_f200w} and \ref{fig:pysersic_residual_f356w}). In the F200W filter, the average absolute residual percentage is $5.01\% \pm 1.71\%$ (median $2.87\% \pm 0.30\%$) with a maximum difference of 47.23\% (for ID = 1090947, which is difficult to model due to its disturbed morphology) and the next highest flux difference of 10.62\%. In the F356W filter, the average absolute residual percentage is $5.62\% \pm 0.65\%$ (median $4.73\% \pm 1.02\%$) with a maximum difference of 18.15\% (again for ID = 1090947) and the next highest flux difference of 9.00\%. Therefore, our imaging shows low central residuals and lacks dominant point sources, ruling out bright luminous AGNs during the current quiescent phase \citep[consistent with][]{Haryana_2025}. However, recent spectroscopic observations at cosmic noon detect gas outflows in recently quenched systems \citep{Belli2024, Park2024, Scholtz2026}. Future JWST NIRSpec IFU spectroscopy will be required to directly measure gas kinematics and outflows in our galaxies for testing these combined mechanisms.

\section{Summary \& Conclusions} \label{sec:conclusions}

We have investigated the structural properties, multi-wavelength size--mass scaling relations, and spatially resolved star formation histories of 26 massive quiescent galaxies at cosmic noon ($2 < z < 3$) in the UDS field using data from the JWST PRIMER \citep{Donnan2024} and MINERVA surveys \citep{Muzzin2025}. We summarize our primary findings below.

\begin{enumerate}

  \item Across the six NIRCam broadband filters, the size--mass relation for massive quiescent galaxies ($\log_{10}(M_*/M_\odot) > 10.3$) exhibits a steep power-law slope of $\alpha \approx 0.51\text{--}0.59$. In the rest-frame optical (F200W), we measure a slope of $\alpha = 0.59 \pm 0.12$ and a characteristic size of $A = 0.83_{-0.07}^{+0.08}\ \mathrm{kpc}$ at pivot mass $M_* = 5 \times 10^{10}\ M_\odot$. In the rest-frame near-infrared (F356W), we derive a consistent slope of $\alpha = 0.56 \pm 0.13$ and an intercept of $A = 0.70_{-0.06}^{+0.07}\ \mathrm{kpc}$. The persistent steep slope ($\alpha \sim 0.6$) across cosmic time demonstrates that the size of massive quiescent galaxies already obeyed a strong mass dependence by cosmic noon, matching low-redshift early-type galaxies \citep{Shen2003} and cosmic noon results \citep{vanderWel2014, Kawinwanichakij2026}.

  \item Galaxy sizes contract systematically toward longer wavelengths, shrinking by $\approx 22\%$ from F115W ($A = 0.87\ \mathrm{kpc}$) to F444W ($A = 0.68\ \mathrm{kpc}$). This size shrinkage corresponds to a shift of $\approx 0.07\ \mathrm{dex}$ in normalization between F200W and F356W. This trend reflects negative radial color gradients, where rest-frame optical imaging captures younger stars at larger radii, while rest-frame near-infrared imaging traces the dense, older stellar mass core \citep{Suess2019, Suess2022}.

  \item In F200W, two-component bulge--disk decompositions reveal that roughly two-thirds ($N = 18$) of the galaxies are bulge-dominated ($B/T \ge 0.5$), while approximately one-third ($N = 8$) retain dominant exponential disks ($B/T < 0.5$). Central bulges are ultra-compact ($R_{e,\mathrm{Bulge}} \approx 0.49\ \mathrm{kpc}$, $n_{\mathrm{bulge}} \approx 3.4\text{--}3.8$) and show negligible size variation with stellar mass across $\log_{10}(M_*/M_\odot) \approx 10.3\text{--}11.6$, in agreement with the mass invariance found in bulges at lower redshifts \citep{Lang2014, Robotham2022}. In contrast, surrounding disks are more extended ($R_{e,\mathrm{Disk}} \approx 1.4\text{--}1.5\ \mathrm{kpc}$), but lie $\approx 0.3\text{--}0.4\ \mathrm{dex}$ below the star-forming galaxy size--mass relations at cosmic noon \citep{vanderWel2014, Ward2024}. Both individual components exhibit minimal size changes between optical and near-infrared wavelengths ($\Delta \log R_e \le 0.04\ \mathrm{dex}$), indicating that the global size contraction arises from the increased flux dominance of the compact bulge at longer wavelengths.

  \item Spatially resolved SED fitting with \pixedfit \citep{Abdurrouf2021} reveals positive radial sSFR gradients across the sample, providing direct empirical evidence for inside-out quenching \citep{Tacchella2016, Haryana_2025}. Specific star formation rates in the outer regions exceed central core values by roughly two orders of magnitude. The radial color profiles show negative gradients in both $U-V$ and $V-J$, confirming the presence of older, redder stellar cores surrounded by younger outskirts.

  \item Spatially resolved formation timescales confirm that the inner regions ($R < 4\ \mathrm{kpc}$) formed earlier ($t_{50} \sim 1.2\ \mathrm{Gyr}$) and quenched more rapidly ($t_q - t_{50} \approx 0.8\ \mathrm{Gyr}$) than the outer regions ($t_{50} \sim 2.1\ \mathrm{Gyr}$ and $t_q - t_{50} \sim 1.2\ \mathrm{Gyr}$), reaching 50\% of their stellar mass at least $\approx 0.7\ \mathrm{Gyr}$ earlier. This chronological sequence supports the dissipative gas compaction framework, where gas inflows at $z > 2$ assemble an ultra-dense stellar core before star formation ceases \citep{Dekel2014, Barro2017}.

  \item The structural properties and quenching patterns are consistent with a lifecycle in which compaction, AGN feedback, and morphological stabilization act sequentially. In this picture, early compaction builds the dense core and channels gas onto the central black hole \citep{Barro2017}, nuclear AGN feedback then heats or expels central gas to drive rapid inside-out quenching \citep{Fabian2012, Nelson2019}, and the massive central bulge may help stabilize the surviving disk against fragmentation \citep{Martig2009, Genzel2014}. The survival of extended disks in our sample agrees with JWST evidence that massive quiescent galaxies at cosmic noon commonly remain rotationally supported \citep{Slob2025}. Directly testing these mechanisms requires gas and kinematic measurements.

\end{enumerate}

Our findings demonstrate that massive quiescent galaxies at cosmic noon were already structurally transformed into compact, bulge-dominated systems while preserving extended disk components. The low central residuals in our imaging rule out luminous active AGNs during the current quiescent phase, consistent with systems observed after the peak blowout phase. Future spectroscopic observations with JWST NIRSpec IFU will directly map the stellar and gas kinematics, measuring potential gas outflows and testing these combined quenching mechanisms in detail.

\begin{acknowledgments}
  We thank the PRIMER and MINERVA collaboration teams for designing the observing programs, developing the data reduction pipelines, and making high-level data products publicly accessible to the astronomical community. We also acknowledge the longstanding community effort to observe the CANDELS and 3D-HST extragalactic deep fields.
  This work is based on observations made with the NASA/ESA/CSA James Webb Space Telescope and the NASA/ESA Hubble Space Telescope. The data were obtained from the Mikulski Archive for Space Telescopes (MAST) at the Space Telescope Science Institute, which is operated by the Association of Universities for Research in Astronomy, Inc., under NASA contract NAS 5-03127 for JWST and NAS 5-26555 for HST. These observations are associated with JWST Cycle 1 Treasury program 1837 (PRIMER) and Cycle 4 Treasury program 7814 (MINERVA) in the Ultra-Deep Survey field, along with archival HST observations from the CANDELS (HST-GO-12064), 3D-HST (HST-GO-12177 and HST-GO-12328), and HST-GO-16872 programs. The spectroscopic redshifts are based on JWST/NIRSpec observations associated with programs 1215 (NIRSpec WIDE), 3543 (EXCELS), 4233 (RUBIES), and 6368 (CAPERS).
  Some of the data products presented herein were retrieved from the Dawn JWST Archive (DJA). DJA is an initiative of the Cosmic Dawn Center (DAWN), which is funded by the Danish National Research Foundation under grant DNRF140.
  VP acknowledges support from the Indian Institute of Science Education and Research (IISER) Pune and the National Centre for Radio Astrophysics (NCRA-TIFR), Pune. YW and PB acknowledge the support of the Department of Atomic Energy, Government of India, under the project title ``Next Generation Instrumentation for Radio Astronomy'', UID number RTI4017.
\end{acknowledgments}

\section{Data Availability}
The modeling outputs and derived parameters from this work are available on Zenodo under DOI \href{https://doi.org/10.5281/zenodo.23669808}{10.5281/zenodo.23669808}. These products include the fitting results and catalogs from \texttt{BAGPIPES}, \texttt{pysersic}, and \texttt{piXedfit} for all 26 galaxies in our sample. The underlying imaging mosaics and photometric catalogs are available through the MINERVA data release, and spectroscopic data are accessible through the DAWN JWST Archive.

\facilities{HST, JWST}

\software{\texttt{astropy} \citep{2013A&A...558A..33A,2018AJ....156..123A,2022ApJ...935..167A},
  \texttt{BAGPIPES} \citep{Carnall2018},
  \texttt{dynesty} \citep{Speagle2020, Koposov2026},
  \texttt{Gnuastro v0.23} \citep{gnuastro,astscript-color-faint-gray},
  \texttt{photutils} \citep{Bradley2025},
  \texttt{piXedfit} \citep{Abdurrouf2021},
  \texttt{PyPHER} \citep{Boucaud2016},
  \texttt{pysersic} \citep{Pasha2023},
  \texttt{scipy} \citep{SciPy2020},
  \texttt{sedpy} \citep{johnson_2021_4582723},
  \texttt{SEP} \citep{Bertin1996, Barbary2016},
  \texttt{stpsf/WebbPSF} \citep{Perrin2012, Perrin2014}
}

\appendix

\section{Multi-Filter Morphological Parameters} \label{sec:appendix_morphology}

In Table~\ref{tab:morph_summary_appendix}, we compile the structural and morphological parameters derived across all six JWST/NIRCam broadband filters (F115W, F150W, F200W, F277W, F356W, and F444W) using \pysersic \citep{Pasha2023}. We report the posterior 50th percentiles (medians) with 16th and 84th percentile uncertainties ($1\sigma$) for single \sersic profile fits (index $n$, projected axis ratio $q \equiv b/a$, and semi-major effective radius $R_e$) and two-component \texttt{sersic\_exp} decompositions (bulge-to-total flux ratio $B/T$, bulge \sersic index $n_{\mathrm{bulge}}$, bulge effective radius $R_{e,\mathrm{Bulge}}$, and disk effective radius $R_{e,\mathrm{Disk}}$).

\begin{deluxetable*}{lcccccccc}[ht!]
  \tabletypesize{\scriptsize}
  \tablecaption{Summary of Multi-Filter Morphological Parameters across JWST NIRCam Bands\label{tab:morph_summary_appendix}}
  \tablehead{
    \colhead{Filter} & \colhead{$\lambda_{\mathrm{rest}}$} & \colhead{S\'ersic $n$} & \colhead{Axis Ratio $q$} & \colhead{$R_e$} & \colhead{$B/T$} & \colhead{$n_{\mathrm{bulge}}$} & \colhead{$R_{e,\mathrm{Bulge}}$} & \colhead{$R_{e,\mathrm{Disk}}$} \\
    \colhead{} & \colhead{($\mu\mathrm{m}$)} & \colhead{} & \colhead{($b/a$)} & \colhead{(kpc)} & \colhead{} & \colhead{} & \colhead{(kpc)} & \colhead{(kpc)}
  }
  \setlength{\tabcolsep}{6pt}
  \startdata
  F115W & $\sim 0.35$ & $2.97_{-0.21}^{+0.21}$ & $0.59_{-0.03}^{+0.06}$ & $0.99_{-0.18}^{+0.12}$ & $0.60_{-0.04}^{+0.05}$ & $4.21_{-0.14}^{+0.07}$ & $0.53_{-0.05}^{+0.08}$ & $1.24_{-0.20}^{+0.35}$ \\
  F150W & $\sim 0.45$ & $2.91_{-0.50}^{+0.42}$ & $0.64_{-0.09}^{+0.08}$ & $0.82_{-0.07}^{+0.23}$ & $0.53_{-0.07}^{+0.07}$ & $3.98_{-0.52}^{+0.12}$ & $0.62_{-0.08}^{+0.06}$ & $1.50_{-0.19}^{+0.18}$ \\
  F200W & $\sim 0.60$ & $2.85_{-0.46}^{+0.37}$ & $0.60_{-0.08}^{+0.07}$ & $0.87_{-0.15}^{+0.19}$ & $0.59_{-0.06}^{+0.04}$ & $3.77_{-0.09}^{+0.44}$ & $0.49_{-0.04}^{+0.11}$ & $1.38_{-0.28}^{+0.20}$ \\
  F277W & $\sim 0.85$ & $2.91_{-0.55}^{+0.38}$ & $0.61_{-0.10}^{+0.12}$ & $0.83_{-0.13}^{+0.20}$ & $0.62_{-0.07}^{+0.05}$ & $3.63_{-0.26}^{+0.43}$ & $0.44_{-0.06}^{+0.06}$ & $1.58_{-0.13}^{+0.22}$ \\
  F356W & $\sim 1.10$ & $2.90_{-0.43}^{+0.32}$ & $0.58_{-0.07}^{+0.09}$ & $0.78_{-0.16}^{+0.16}$ & $0.57_{-0.05}^{+0.05}$ & $3.37_{-0.37}^{+0.21}$ & $0.49_{-0.09}^{+0.05}$ & $1.54_{-0.21}^{+0.17}$ \\
  F444W & $\sim 1.35$ & $3.02_{-0.49}^{+0.23}$ & $0.57_{-0.07}^{+0.09}$ & $0.76_{-0.16}^{+0.14}$ & $0.69_{-0.15}^{+0.03}$ & $3.46_{-0.27}^{+0.09}$ & $0.47_{-0.11}^{+0.06}$ & $1.56_{-0.18}^{+0.07}$ \\
  \enddata
  \tablecomments{Median structural parameters and bootstrapped 68\% confidence intervals (1000 iterations) for our sample of 26 massive quiescent galaxies at $z \sim 2.0$--$2.9$. Columns report: NIRCam filter; approximate rest-frame wavelength $\lambda_{\mathrm{rest}}$; single \sersic index $n$; projected minor-to-major axis ratio $q \equiv b/a$; semi-major effective half-light radius $R_e$; bulge-to-total flux ratio $B/T$ from two-component \texttt{sersic\_exp}; bulge \sersic index $n_{\mathrm{bulge}}$; bulge effective radius $R_{e,\mathrm{Bulge}}$; and exponential disk effective radius $R_{e,\mathrm{Disk}}$.}
\end{deluxetable*}

\section{Forward-Modeling JWST/NIRCam PSFs and Comparison with Empirical Stars} \label{sec:appendix_psf}

Parametric \sersic modeling and multi-band PSF homogenization require accurate point spread function models. While synthetic optical models from \texttt{WebbPSF} \citep{Perrin2012, Perrin2014} capture nominal diffraction optics, observational studies show that simulated PSFs are $1\%$--$2\%$ narrower than real point sources in drizzled mosaics \citep[e.g.,][Appendix~B]{Weaver2024}. To evaluate this broadening across the MINERVA and PRIMER imaging in the UDS field, we forward-drizzle optical continuous PSFs generated using \texttt{stpsf} with epoch-specific MAST Optical Path Difference (OPD) maps for PRIMER (Cycle~1) and MINERVA (Cycles~2 \& 3).

Following \citet{Casertano2000} and \citet{Fruchter2002}, we express the forward-modeled mosaic PSF ($\mathrm{PSF}_{\mathrm{mosaic}}$) as
\begin{equation}
  \begin{split}
    \mathrm{PSF}_{\mathrm{mosaic}}(\mathbf{x}) = & \ \mathrm{PSF}_{\mathrm{opt}}(\mathbf{x}) * G_{\mathrm{jitter}}(\mathbf{x})                          \\
                                                 & * \Pi_{\mathrm{det}}(\mathbf{x}) * \Pi_{\mathrm{drop}}(\mathbf{x}) * \Pi_{\mathrm{out}}(\mathbf{x}),
  \end{split}
  \label{eq:forward_psf}
\end{equation}
where $*$ denotes 2D spatial convolution. Here, $\mathrm{PSF}_{\mathrm{opt}}$ is the continuous optical PSF ($4\times$ oversampled) generated from the complex pupil function incorporating the measured telescope OPD; $G_{\mathrm{jitter}}$ is Gaussian pointing jitter ($1\sigma = 7\ \mathrm{mas}$); $\Pi_{\mathrm{det}}$ is native detector pixel integration ($0\farcs0311$ for short wavelength filters; $0\farcs0630$ for long wavelength filters); $\Pi_{\mathrm{drop}}$ is the Drizzle drop footprint ($w_{\mathrm{drop}} = 0.75 \cdot s_{\mathrm{in}}$, matching $\texttt{pixfrac} = 0.75$ in the science image header); and $\Pi_{\mathrm{out}}$ integrates onto the output mosaic grid ($0\farcs040\ \mathrm{pixel}^{-1}$).

\begin{figure*}[ht!]
  \centering
  \includegraphics[width=\textwidth]{ 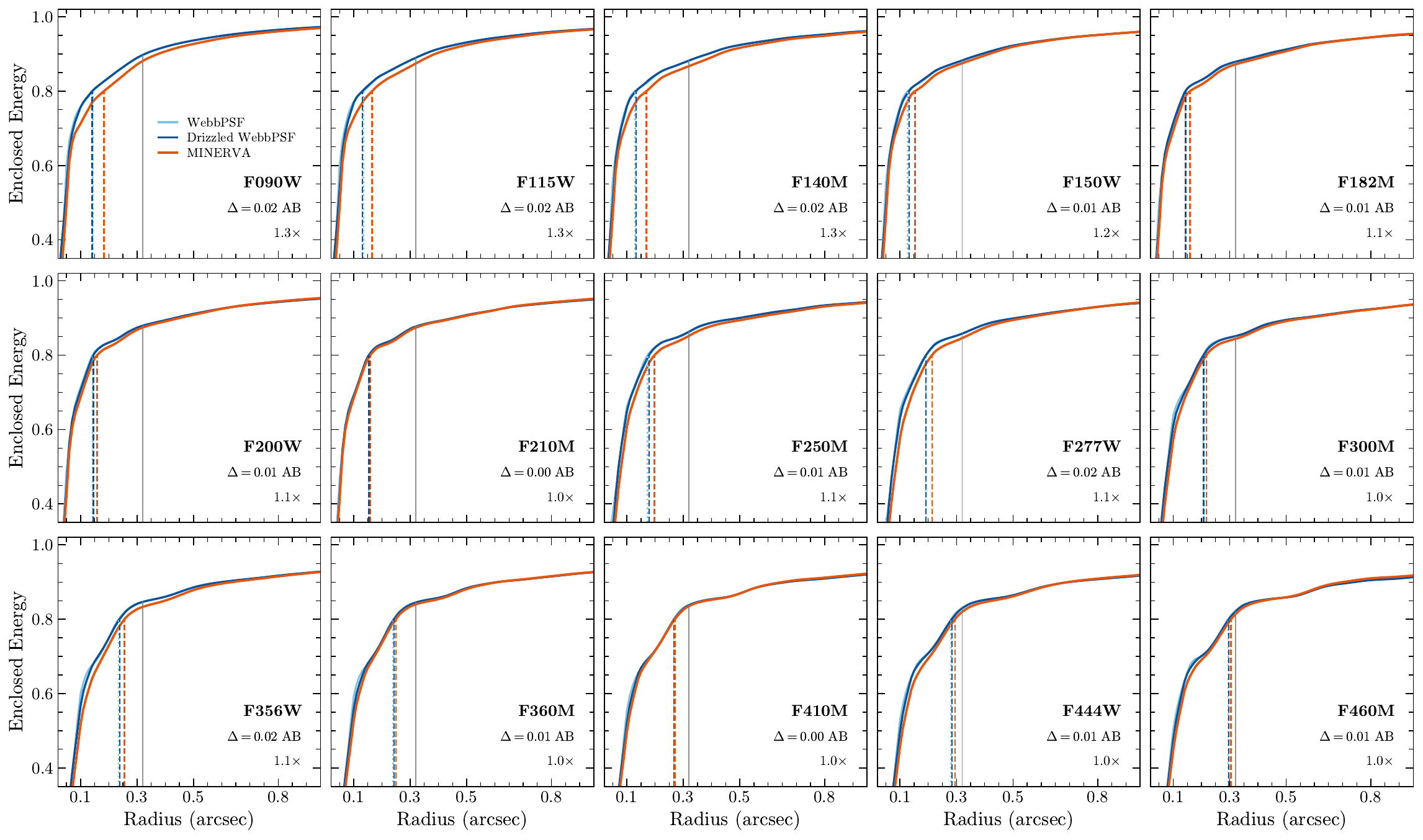}
  \caption{Circular aperture enclosed energy profiles for all 15 JWST/NIRCam filters in the MINERVA mosaics, following \citet[][Figure~17]{Weaver2024}. In each panel, light-blue curves display the native detector-sampled \texttt{WebbPSF} model, dark-blue curves show the forward-drizzled model incorporating pointing jitter ($1\sigma = 7\ \mathrm{mas}$), native detector integration, drizzle drop deposition, and output grid integration (Equation~\ref{eq:forward_psf}), and orange curves show stacked empirical stars from MINERVA. Vertical dashed lines mark the 80\% enclosed energy radius ($r_{80}$) for each model and empirical PSF, while the thin vertical line indicates the fiducial photometric aperture radius ($r = 0\farcs32$). Annotations list the photometric difference $\Delta\mathrm{AB}$ at $r = 0\farcs32$ and the empirical-to-drizzled $r_{80}$ ratio. Curves are normalized to the total exit-pupil flux using STScI encircled energy calibration tables, correctly reflecting diffraction wing losses outside the $4\farcs04$ cutout.}
  \label{fig:psf_enclosed_energy}
\end{figure*}

Following \citet[][Appendix~B]{Weaver2024}, we normalize radial enclosed energy (EE) profiles to the telescope exit pupil ($r \sim 4\farcs0$) using STScI calibration tables. Figure~\ref{fig:psf_enclosed_energy} shows the resulting curves of growth as the energy enclosed as a function of circular aperture radius across all 15 NIRCam filters for native \texttt{WebbPSF} (light blue), forward-drizzled \texttt{WebbPSF} (dark blue), and stacked empirical stars from MINERVA (orange). While forward-drizzling accounts for most of the broadening relative to native \texttt{WebbPSF}, the empirical PSFs are still slightly wider at fixed EE (e.g., $80\%$ enclosed energy radius, $r_{80}$). These profile differences translate into subtle aperture photometric offsets of $|\Delta\mathrm{AB}| \le 0.01$--$0.02\ \mathrm{mag}$ (median $0.011\ \mathrm{mag}$) at the fiducial aperture radius ($r = 0\farcs32$). Following \citet{Weaver2024}, we therefore adopt empirical PSFs constructed directly from science mosaics for our morphological fitting and multi-band PSF homogenization.

\clearpage
\bibliographystyle{aasjournalv8}
\bibliography{references}

\end{document}